\documentclass[10pt,conference]{IEEEtran}
\IEEEoverridecommandlockouts

\usepackage[ruled,vlined,linesnumbered]{algorithm2e}
\usepackage{xcolor}
\usepackage{subcaption}
\usepackage{booktabs}

\newcommand{\envi}{\ensuremath{\mathit{Env}}}
\newcommand{\cont}{\ensuremath{\mathit{Con}}}
\newcommand{\wpr}{\ensuremath{\mathit{CP}}}

\newcommand{\ltltba}{\mathit{LTL2BA}}
\newcommand{\hpsi}{\mathcal{H}_\psi}
\newcommand{\apsi}{A_\psi}
\newcommand{\hnpsi}{\mathcal{H}_{\neg\psi}}
\newcommand{\anpsi}{A_{\neg\psi}}
\newcommand{\aut}{\mathit{Aut}}
\newcommand{\fin}{\mathit{final}}
\newcommand{\succc}{\mathit{Succ}}
\newcommand{\destp}{\mathit{DestPair}}

\newcommand{\qed}{\ensuremath{\Box}}

\newcommand {\reals} {\ensuremath{\mathbb{R}}}

\newcommand {\ints} {\ensuremath{\mathbb{Z}}}

\newcommand {\union} {\cup}

\newcommand {\myOr} {\ensuremath{\vee}}
\newcommand {\myAnd} {\ensuremath{\wedge}}

\newcommand {\limplies} {\ensuremath{\rightarrow}}

\renewcommand {\models} {\ensuremath{\vDash}}

\newcommand {\emptyword} {\ensuremath{\epsilon}}
\newcommand {\true} {\ensuremath{\mathit{true}}}
\newcommand {\false} {\ensuremath{\mathit{false}}}

\newcommand {\A}        {\ensuremath{\mathcal{A}}}

\newcommand {\T}        {\ensuremath{\mathcal{T}}}

\newcommand {\SA}        {\ensuremath{\mathcal{S}}}

\newcommand {\ltl} {\ensuremath{\mathit{LTL}}}
\newcommand {\Until}        {\ensuremath{U}}
\newcommand {\X}        {\ensuremath{X}}
\newcommand {\F}        {\ensuremath{F}}
\newcommand {\Gl}        {\ensuremath{G}}

\newcommand {\constr}[1] {\ensuremath{\mathit{Constr}(#1)}}
\newcommand {\val}[2] {\ensuremath{\mathbf{#1}_{#2}}}
\newcommand {\myGame} {\ensuremath{\mathcal{G}}}
\newcommand {\init} {\ensuremath{\mathit{Init}}}
\newcommand {\winreg}[2] {\ensuremath{\mathit{winreg}_{#1}(#2)}}
\newcommand {\myprod} {\ensuremath{\!\otimes\!}}
\newcommand {\gensysltl} {\textsc{GenSys-LTL}}

\newcommand {\subseteqfin} {\ensuremath{\subseteq_\mathit{fin}}}
\newcommand {\finrun} {\ensuremath{\mathit{run}}}
\newcommand {\move} {\ensuremath{\mathit{mov}}}
\newcommand {\qelim} {\ensuremath{\mathit{QElim}}}

\usepackage{longtable}
\usepackage{supertabular}

\usepackage{cite}
\usepackage{amsmath,amssymb,amsfonts}
\usepackage{algorithmic}
\usepackage{graphicx}
\usepackage{textcomp}
\usepackage{xcolor}
\usepackage{url}
\def\BibTeX{{\rm B\kern-.05em{\sc i\kern-.025em b}\kern-.08em
    T\kern-.1667em\lower.7ex\hbox{E}\kern-.125emX}}

\newtheorem{example}{Example}[section]
\newtheorem{theorem}{Theorem}

\begin{document}

\title{Towards Efficient Controller Synthesis Techniques for Logical LTL Games}

\author{\IEEEauthorblockN{Stanly Samuel}
\IEEEauthorblockA{\textit{Computer Science and Automation} \\
\textit{Indian Institute of Science}\\
Bengaluru, India \\
stanly@iisc.ac.in}
\and
\IEEEauthorblockN{Deepak D’Souza}
\IEEEauthorblockA{\textit{Computer Science and Automation} \\
\textit{Indian Institute of Science}\\
Bengaluru, India \\
deepakd@iisc.ac.in}
\and
\IEEEauthorblockN{Raghavan Komondoor}
\IEEEauthorblockA{\textit{Computer Science and Automation} \\
\textit{Indian Institute of Science}\\
Bengaluru, India \\
raghavan@iisc.ac.in}
}

\maketitle

\begin{abstract}
Two-player games are a fruitful way to represent and reason about
several important synthesis tasks.
These tasks include controller synthesis (where one asks for a controller for a
given plant such that the controlled plant satisfies a given temporal
specification), program repair (setting values of variables to avoid
exceptions), and synchronization synthesis (adding lock/unlock
statements in multi-threaded programs to satisfy safety assertions).
In all these applications, a solution directly corresponds to a
winning strategy for one of the players in the induced game.
In turn, \emph{logically-specified} games offer a powerful way to model these
tasks for large or infinite-state systems. Much
of the techniques proposed for solving such games typically
rely on abstraction-refinement or template-based solutions.
In this paper, we show how to apply classical fixpoint algorithms, that
have hitherto been used in explicit, finite-state, settings, to a
symbolic logical setting.
We implement our techniques in a tool called \gensysltl\ and show that
they are not only effective in synthesizing
valid controllers for a variety of challenging benchmarks from the
literature, but often compute maximal winning regions and
maximally-permissive controllers.  We achieve \textbf{46.38X speed-up} over the state of the art and also scale well for non-trivial LTL specifications.

\end{abstract}

\begin{IEEEkeywords}
reactive synthesis,  symbolic algorithms,  program synthesis,  program
repair,  two-player games 
\end{IEEEkeywords}

\section{Introduction}
\label{sec:intro}

Two-player games are games played between two players called the Controller and
the Environment, on a game graph or arena.
The players generate an infinite sequence of states (a so-called
``play'') in the game by making moves alternately, from a specified
set of legal moves.
The Controller wins the play if the sequence of states satisfies a winning
condition (e.g.,  a Linear-Time Temporal Logic (LTL) formula).
The central question in these games is whether a player (typically the
Controller) has a winning strategy from a given set of initial states
(called the realizability problem), or more generally, to compute the
set of states from which she wins (i.e.\@ the winning region).

Games are a fruitful way to model and reason about several important
problems in Sofware Engineering, like \emph{controller synthesis}
\cite{AsarinMP94} (where a
winning strategy for the Controller in the associated game directly
corresponds to a valid controller for the system);
\emph{program repair} \cite{JobstmannGB05} (strategy corresponds to
corrected program); \emph{synchronization synthesis}
\cite{VechevYY10} (strategy corresponds to
appropriate placement of synchronization statements in a concurrent
program); and \emph{safety shield synthesis} \cite{ZhuXMJ19} (winning region
corresponds to region in which the neural-network
based controller is allowed to operate without the shield stepping in).

Classical techniques for solving games
\cite{Buchi1969,Thomas1995,MalerPS95}, and more recent improvements
\cite{KupfermanV05,jobstmann2018graph,filiot2011antichains}, work on
finite-state games, by iteratively computing sets of states till a
fixpoint is reached.
These algorithms typically allow us to compute the exact winning
region and thereby answer the realizability question as well.

In recent years, \emph{logical games} -- where the moves of the players
are specified by logical formulas on the state variables -- have
attracted much attention, due to their ability to model large or
infinite-state systems.
Techniques proposed for these games range from constraint solving
\cite{consynth}, finite unrollings and generalization
\cite{FarzanK18}, 
CEGAR-based abstraction-refinement
\cite{Finkbeiner0PS19,FinkbeinerHP22,raboniel},
counterexample-based learning \cite{dtsynth},
combination of Sygus
and classical LTL synthesis \cite{temos}, and
solver-based enumeration \cite{WuQCRFLDX20}.
Among these Beyene et al \cite{consynth} address general LTL specs, while the
others handle only safety or reachability specs.
Furthermore, none of these techniques are able to compute precise
winning regions.

In this paper we show that symbolic fixpoint techniques can be
effectively applied to solve logical games with general LTL specifications.
We propose a bouquet of techniques that target different classes of
LTL specs, from simple specs which directly involve a safety, reachability,
B\"uchi, or Co-B\"uchi condition on the states of the game, to
those for which the formula automata are non-deterministic.
The techniques we propose are guaranteed (whenever they terminate) to
compute the \emph{exact} winning region, and, for certain kinds of
games, output a finite-memory winning strategy as well.

We show how to implement these algorithms in a logical setting, by
leveraging the right tactics in available SMT solvers.
We evaluate our prototype tool, called \gensysltl, on a host of
benchmarks from the literature.
Our tool terminates on all benchmarks except one, and takes an average
time of 7.1\,sec to solve each benchmark.
It thus outperforms the state-of-the-art tools in terms of the number
of instances solved, and by an order of magnitude in terms of running
time.


\section{Preliminaries}
\label{sec:prelim}

We will be dealing with standard first-order logic of addition ($+$),
comparison ($<$), and constants $0$ and $1$, interpreted over
the domain of reals
$\reals$ (or a subset of $\reals$ like the integers $\ints$).
The atomic formulas in this logic are thus of the form
$a_1x_1 + \cdots + a_nx_n \sim c$, where $a_i$s and $c$ are integers, $x_i$s are
variables, and ``$\sim$'' is a comparison symbol in $\{<,
\leq, =, \geq, >\}$.
We will refer to such formulas as \emph{atomic constraints}, and to
boolean combinations of such formulas (or equivalently,
quantifier-free formulas) as \emph{constraints}.
We will denote the set of constraints over a set of variables $V$ by
$\constr{V}$. 

For a set of variables $V$, a $V$-\emph{valuation} (or a
$V$-\emph{state}) is 
simply a mapping $s: V \rightarrow \reals$.
Given a constraint $\delta$ over a set of variables $V$, and a
$V$-state $s$, we say $s$ \emph{satisfies} $\delta$, written $s \models
\delta$, if the constraint $\delta$ evaluates to true in $s$ (defined
in the expected way).
We denote the set of $V$-states by $\val{V}{\reals}$.
A \emph{domain mapping} for $V$ is a map $D : V \rightarrow
2^{\reals}$, which assigns a domain $D(x) \subseteq \reals$
for each variable $x$ in $V$.
We will call a $V$-state $s$ whose range respects
a domain mapping $D$, in that \@ for each $x \in V$, $s(x) \in D(x)$),
a $(V,D)$-state, and
denote the set of such $(V,D)$-states by $\val{V}{D}$.  We also denote the cardinality of a set $S$ as $|S|$.

We will sometimes write $\varphi(X)$ to denote that the free variables
in a formula $\varphi$ are among the variables in the set $X$.
For a set of variables $X = \{x_1, \ldots, x_n\}$ we will sometimes
use the notation $X'$ to refer to the set of ``primed'' variables
$\{x_1', \ldots, x_n'\}$.
For a constraint $\varphi$ over a set of variables $X = \{x_1, \ldots,
x_n \}$, we will write $\varphi[X'/X]$ (or simply $\varphi(X')$ when
$X$ is clear from the context) to represent the constraint
obtained by substituting $x_i'$ for each $x_i$ in $\varphi$.

Finally, we will make use of standard notation from formal
languages.
For a (possibly infinite) set $S$, we will view finite and infinite
sequences of elements of $S$ as finite or infinite \emph{words} over
$S$.
We denote the empty word by $\emptyword$.
If $v$ and $w$ are finite words and $\alpha$ an infnite word over $S$,
we denote the conatenation of $v$ and $w$ by $v \cdot w$, and the
concatenation of $v$ and $\alpha$ by $v \cdot \alpha$.
We will use $S^*$ and $S^\omega$ to denote, respectively, the
set of finite and infinite words over $S$.


\section{LTL and Automata}
\label{sec:ltl}

We will make use of a version of Linear-Time Temporal Logic (LTL)
\cite{Pnueli77} where propositions are atomic constraints over a set
of variables $V$ (as in Holzmann \cite{spin2004}, for example).

Let $V$ be a set of variables.
Then the formulas of $\ltl(V)$ are given by:
\[
\psi ::= \delta \ | \ \neg \psi \ | \ \psi \vee \psi \ | \ \X \psi \ | \ \psi \Until
\psi,
\]
where $\delta$ is an atomic constraint over $V$.
The formulas of $\ltl(V)$ are intepreted over an infinite sequence of
$V$-states.
For an $\ltl(V)$ formula $\psi$ and an infinite sequence of
$V$-states $\pi = s_0 s_1 \cdots$, we define when $\psi$ is
satisfied at position $i$ in $\pi$, written $\pi, i \models \psi$,
inductively as follows:
\[
\begin{array}{lll}
  \pi, i \models \delta & \mbox{iff} & s_i \models \delta \\
  \pi, i \models \neg \psi & \mbox{iff} & \pi, i  \not\models \psi \\
  \pi, i \models \psi \vee \psi' & \mbox{iff} & \pi, i  \models \psi
  \mbox{ or } \pi, i \models \psi' \\
  \pi, i \models \X \psi   & \mbox{iff} & \pi, i+1  \models \psi \\
  \pi, i \models \psi \Until \psi' & \mbox{iff} & \exists k \geq
  i \mbox{ s.t.\@ } \pi, k  \models \psi' \mbox { and } \\
  &            & \forall j\!: i \leq j < k \limplies \pi, j \models \psi. 
\end{array}
\]
We say $\pi$ \emph{satisfies} $\psi$, written $\pi \models \psi$, if $\pi, 0
\models \psi$.
We will freely make use of the derived operators $\F$ (``future'') and
$\Gl$ (``globally'') defined by $\F \psi \equiv \true \,\Until\, \psi$ and $\Gl
\psi \equiv \neg \F \neg \psi$, apart from the boolean operators
$\wedge$ (``and''), $\limplies$ (``implies''), etc.

An \emph{$\omega$-automaton} \cite{Thomas1990} $\A$ over a set of variables $V$, is a
tuple $(Q, I, \T, F)$ where
$Q$ is a finite set of states, $I \subseteq Q$ is a set of initial
states, $\T \subseteqfin Q \times \constr{V} \times Q$ is a
``logical'' transition relation, and $F \subseteq Q$ is a set
of final states.
The logical transition relation $\T$ induces a concrete transition
relation $\Delta_{\T} \subseteq Q \times \val{V}{\reals} \times Q$,
given by $(q,s,q')
\in \Delta_{\T}$ iff there exists $(q,\delta,q') \in \T$ such that $s
\models \delta$. A \emph{run} of $\A$ on an infinite sequence of $V$-states
$\pi = s_0s_1 \cdots$ is an infinite sequence of states
$\rho = q_0 q_1 \cdots$, such that $q_0 \in I$, and for each $i$, $(q_i, s_i,
q_{i+1}) \in \Delta_{\T}$.

We say an $\omega$-automaton $\A = (Q, I, \T, F)$ over $V$, is \emph{deterministic}
if $I$ is singleton, and for every $q \in Q$ and $V$-state $s$, there
is at most one $q' \in 
Q$ such that $(q, s, q') \in \Delta_\T$.
Similarly we say $\A$ is \emph{complete} if for every $q \in Q$ and
$V$-state $s$, there exists a $q' \in Q$ such that $(q, s, q') \in
\Delta_\T$.

An $\omega$-automaton can be viewed as either a \textit{B\"uchi} \cite{buchi1990},  \textit{Co-B\"uchi},  \textit{Universal Co-B\"uchi}, or \textit{Safety} automaton based on how the \emph{runs} for a given $V$-state sequence $\pi$ are \textit{accepted} using the final states $F$,  described as follows.
A run $\rho = q_0q_1\cdots$ of  $\A $ is \emph{accepting} by the
\textit{B\"uchi} acceptance condition if for infinitely many $i$, we
have $q_i \in F$, and a $V$-state sequence $\pi$ is accepted by $\A$ if
there exists such a run $\rho$ for $\pi$.  A \emph{B\"uchi Automaton} is an $\omega$-automaton where $F$ is viewed as a B\"uchi acceptance condition.
Similarly, by the \textit{Co-B\"uchi} acceptance condition $\rho$ is \emph{accepting} if it visits $F$ only a finite number of times and a $V$-state sequence $\pi$ is accepted by $\A$ if there
exists such a run $\rho$ for $\pi$.
We call such an automaton a \emph{Co-B\"uchi Automaton (CA)}.
The \textit{Universal Co-B\"uchi} acceptance condition states that a
run $\rho$ of  $\A $ is \emph{accepting} if it visits $F$ only a
finite number of times, and a $V$-state sequence $\pi$ is accepted by
$\A$ if \emph{all} runs $\rho$ for $\pi$ are accepting.
We call such an automaton a \emph{Universal Co-B\"uchi Automaton
  (UCA)}.
%
%
Finally, we can view an $\omega$-automaton $\A$ as a \emph{safety} automaton, by saying that
$\A$ accepts $\pi$ iff there is a run of $\A$ on $\pi$ which never
visits a state outside $F$.
We denote by $L(\A)$ the set of $V$-state sequences accepted by an
$\omega$-automaton $\A$.

It is well-known that any LTL formula $\psi$ can be translated into
a (possibly non-deterministic) B\"uchi automaton $\A_\psi$ that
accepts precisely the models of 
$\psi$ \cite{WolperVS83}.
The same construction works for $\ltl(V)$ formulas, by treating each
atomic constraint as a propositional variable.
Henceforth, for an $\ltl(V)$ formula $\psi$ we will denote the
corresponding formula automaton by $\A_{\psi}$.

\begin{figure}
  \centering
  \includegraphics[scale=0.35]{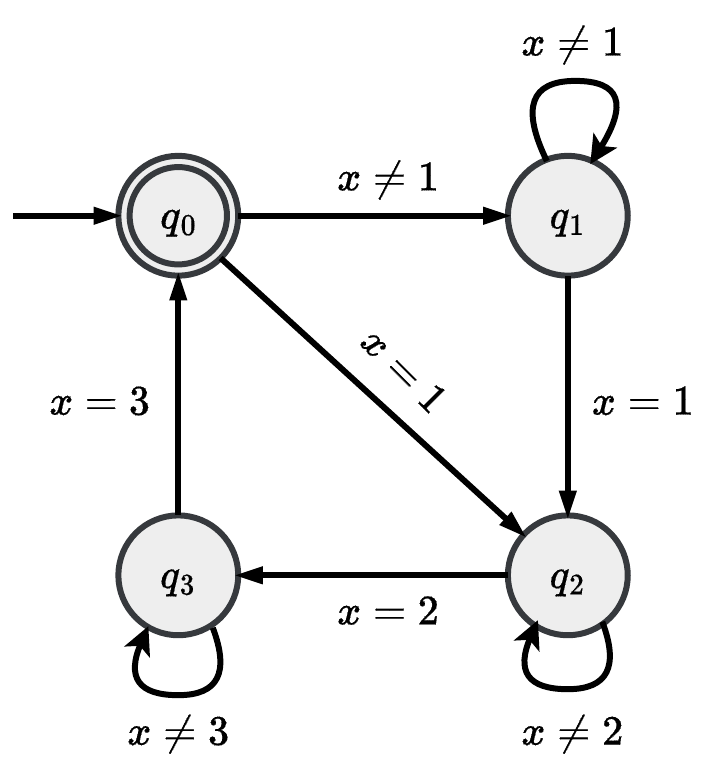}
  \caption{B\"uchi automaton $A_\psi$ for the LTL formula $\psi = \Gl
  (\F(x=1) \myAnd \F(x=2) \myAnd \F(x=3))$. Final states are indicated
with double circles.}
  \label{fig:BA-inf-1-2-3}
\end{figure}

Fig.~\ref{fig:BA-inf-1-2-3} shows a formula automaton $\A_{\psi}$ for
the $\ltl(V)$ formula $\psi = \Gl(F(x=1)
\myAnd F(x=2) \myAnd F(x=3))$ from Example~\ref{example:elevator},
where $V = \{x\}$. The automaton can be seen to be deterministic.

\section{LTL Games}
\label{sec:games}

In this section we introduce our notion of logically specified games,
where moves are specified by logical constraints and winning
conditions by LTL formulas. These games are similar to the formulation
in Beyene et al \cite{consynth}.

A \emph{2-player logical game with an LTL winning condition} (or
simply an \emph{LTL game}) is of the form
\[
\myGame = (V, D, \cont, \envi, \psi), \ \ \ where
\]
\begin{itemize}
\item $V$ is a finite set of variables.
\item $D: V \rightarrow 2^\reals$ is a domain mapping for $V$.
  \item $\cont$ and $\envi$ are both constraints over $V \union V'$,
    representing the moves of Player~$C$ and Player~$E$ respectively.
  \item $\psi$ is an $\ltl(V)$ formula.
\end{itemize}

The constraint $\cont$ induces a transition relation
\[
\Delta_\cont
\subseteq \val{V}{D} \times \val{V}{D}
\]
given by $(s,s') \in \Delta_\cont$
iff $s$ and $s'$ are $(V,D)$-states, and \mbox{$(s,s') \models
  \cont$}.
We use the notation $(s,s') \models \cont$ to denote the fact that
$t_{s,s'} \models \cont$, where $t_{s,s'}$ is the valuation over $V
\union V'$ which maps each
\mbox{$x \in V$} to $s(x)$ and $x' \in V'$ to $s'(x)$.
In a similar way, $\envi$ induces a transition relation
$\Delta_\envi \subseteq \val{V}{D} \times \val{V}{D}$.
For convenience we will assume that the $C$-moves are ``complete'' in
that for every $(V,D)$-state $s$,
there is a $(V,D)$-state $s'$ such that $(s,s') \in \Delta_\cont$;
and similarly for Player~$E$.

A play in $\myGame$ is an sequence of $(V,D)$-states obtained by an alternating
sequence of moves of Players~$C$ and $E$, with Player~$C$ making the
first move.
More precisely, an \emph{(infinite) play} of $\myGame$, starting from a $(V,D)$-state $s$,
is an infinite sequence of $(V,D)$-states $\pi = s_0 s_1 \cdots$, such
that
\begin{itemize}
\item $s_0 = s$, and
\item for all $i$, $(s_{2i}, s_{2i+1}) \in \Delta_\cont$ and
  $(s_{2i+1}, s_{2i+2}) \in \Delta_\envi$.
\end{itemize}
We similarly define the notion of a \emph{finite} play $w$ in the
expected manner.
We say a play $\pi$ is \emph{winning} for Player~$C$ if it satisfies
$\psi$ (i.e.\@ $\pi \models \psi$); otherwise it is winning for
Player~$E$.

A strategy for Player~$C$ assigns to odd-length sequences of
states, a non-empty subset of states that correspond to legal moves of $C$.
More precisely, a \emph{strategy} for Player~$C$ in $\myGame$ is a
partial map
\[
\sigma: ((\val{V}{D} \cdot \val{V}{D})^* \cdot \val{V}{D}) \rightharpoonup
2^{\val{V}{D}}
\]
satisfying the following constraints.
We first define when 
a finite play $w$ is \emph{according to} 
$\sigma$, inductively as follows:
\begin{itemize}
  \item $s$ is according to $\sigma$
    iff $s$ belongs to the domain of $\sigma$.
  \item if $w \cdot s$ is of odd length and according to $\sigma$, and
    $s' \in \sigma(w \cdot s)$, then 
    $w \cdot s \cdot s'$ is according to $\sigma$.
  \item if $w\cdot s$ is of even length and according to $\sigma$, and
    $(s,s') \in \Delta_{\envi}$, then $w \cdot s \cdot s'$ is
    according to $\sigma$.
\end{itemize}
For $\sigma$ to be a valid strategy in $\myGame$, we require
that for every finite play $w\cdot s$ of odd-length,
which is according to $\sigma$, $\sigma(w\cdot s)$ must be
defined and non-empty, and for each $s' \in \sigma(w \cdot s)$ we must
have $(s,s') \in \Delta_{\cont}$.

Finally, a strategy $\sigma$ for Player~$C$ is \emph{winning} from a
$(V,D)$-state $s$ in its domain, if
every play that starts from $s$ and is according to $\sigma$, is
winning for Player~$C$ (i.e.\@ the play satisfies $\psi$).
We say $\sigma$ itself is \emph{winning} if it is winning from every
state in its domain.
We say Player~$C$ \emph{wins} from a $(V,D)$-state $s$ if it has a
strategy which is wining from $s$.
We call the set of $(V,D)$-states from which Player-$C$ wins, the
\emph{winning region} for Player~$C$ in $\myGame$, and denote it
$\winreg{C}{\myGame}$.
The analogous notions for Player~$E$ are defined similarly.

We close this section with some further notions about strategies.
We say that a winning strategy $\sigma$ for $C$ is \emph{maximal} if
its domain is 
$\winreg{C}{\myGame}$, and for every strategy
$\sigma'$ for $C$ that is winning from a state $s$ in
$\winreg{C}{\myGame}$, we have $\sigma'(w) \subseteq \sigma(w)$
for each odd-length play $w$ from $s$ according to $\sigma'$.
A strategy $\sigma$ for $C$ is called a 
\emph{(finitely-representable) finite memory} strategy,
if it can be represented by a ``Mealy-style'' \emph{strategy automaton} (see
Fig.~\ref{fig:strategy-elevator}).
This is a finite-state automaton similar to a deterministic B\"uchi
automaton, but
with a partition of the states into controller and environment
states. The initial states are environment states.
The states in the domain of the
strategy are those that satisfy one of the outgoing guards from the
initial state.
Each controller state $q$ has a label $\move(q)$ associated with
it in the form of a constraint over $V \union V'$, which denotes a
subset of moves allowed by $\cont$.
The automaton represents a strategy $\sigma$ in which $\sigma(w)$ for
odd-length $w$ is given by the label 
$\move(q)$ of the state $q$ reached by the automaton on reading $w$.
Finally, a \emph{memoryless} strategy is one that is represented by a
strategy automaton with a \emph{single} environment state.

Synthesizing winning strategies will be easier when the controller's
moves are \emph{finitely non-deterministic}, in that $\cont$ is given by
a disjunction $\cont_1 \myOr \cdots \myOr \cont_k$, where each constraint
$\cont_i$ is \emph{deterministic} (in that whenever $(s,s') \models \cont_i$
and $(s,s'') \models \cont_i$, we have $s' = s'')$.
We call such a game a \emph{finitely non-deterministic} (FND) game.

We illustrate some of these notions through an example below adapted from
\cite{raboniel}.

\begin{example}[Elevator]
\label{example:elevator}
Consider a game $\myGame_1$ which models an elevator control problem,
where the system's state is represented by a single variable $x$ of
type integer, indicating the floor the elevator is currently
positioned at. The controller can choose to move the elevator up or down by one
floor, or stay at the same floor. The environment
does nothing (simply ``skips'').
The specification requires us
visit each of Floor~1, 2, and 3
infinitely often.
The game $\myGame_1$ has the following components: the set of
variables $V$ is $\{x \}$, and the domain map $D$ is given by $D(x) =
\ints$.
The moves of Player~$C$ and Player~$E$ are given by the constraints
$\cont$: $x' = x \myOr x' = x+1 \myOr x'=x-1$,
and $\envi$: $x'= x$, respectively.
The LTL specification $\psi$ is
\(
\Gl(F(x=1) \myAnd F(x=2) \myAnd F(x=3)).
\)
The game is easily seen to be finitely non-deterministic.

The ``game graph'' is shown in Fig.~\ref{fig:game-graph-elevator}. For
convenience we visualize the game as having two copies of the state
space, one where it is the turn of Player~$C$ to make a move
(denoted by circle states on the left) and the other where it is
Player~$E$'s turn to move (indicated by square states on the
right). The moves of $C$ go from left to right, while those of $E$ go
from right to left.

\begin{figure}
  \centering
  \begin{subfigure}[b]{0.28\linewidth}
  \centering
  \includegraphics[scale=0.3]{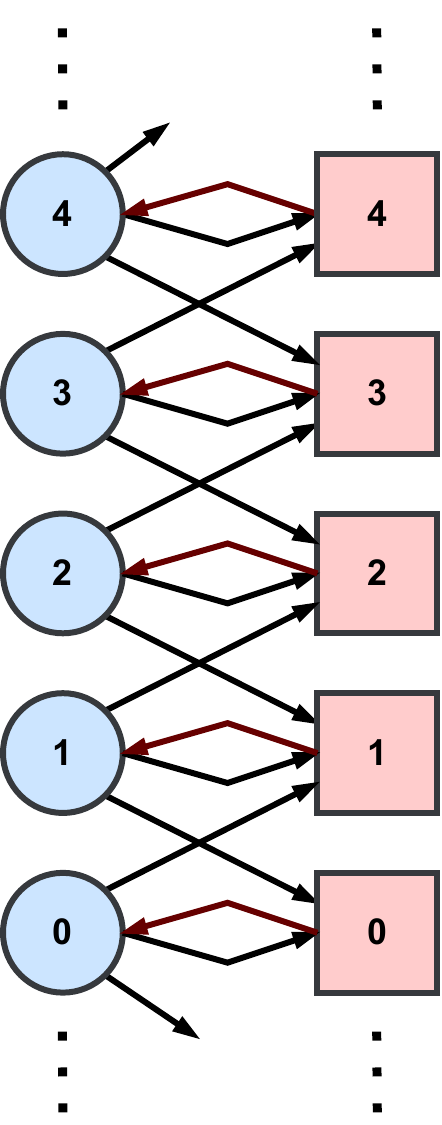}
 \caption{Game graph}
  \label{fig:game-graph-elevator}
  \end{subfigure}
  \begin{subfigure}[b]{0.70\linewidth}
  \centering
\includegraphics[scale=0.35]{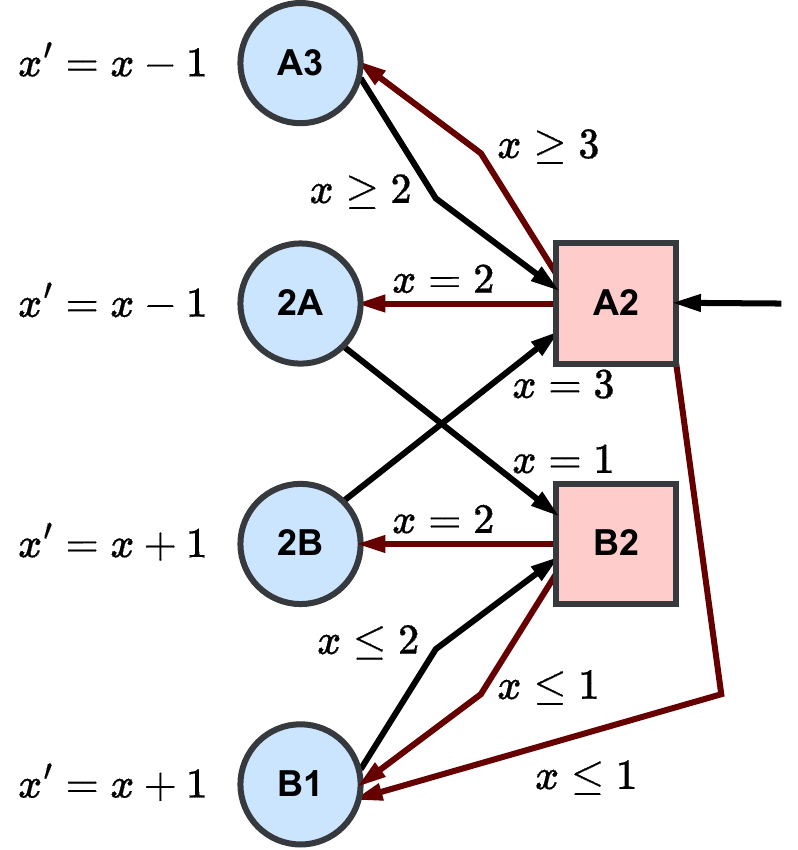}
  \caption{Strategy automaton}
  \label{fig:strategy-elevator}
  \end{subfigure}
\caption{Game graph and strategy for $C$ in Elevator game}
\end{figure}


Player~$C$ has a winning strategy from all states;
for example, by playing $x' = x -
1$ from Floor~3 and above; $x' = x+1$ from Floor~1 and below;
and $x' = x + 1$ and $x' = x - 1$ from Floor~2,
depending on whether it was last in
Floor~1 or 3 respectively.
This finite-memory strategy is shown by the strategy automaton in
Fig.~\ref{fig:strategy-elevator}.



It is easy to see that a memoryless winning strategy does
\emph{not} exist for Player~$C$, as
it cannot afford to play the \emph{same} move from state $x\mapsto 2$ (it
must keep track of the direction in which the lift
is coming from). \hfill \qed
\end{example}

We close this section with a description
of the problems we address in this
paper.
The main problems we address are the following:
\begin{enumerate}
\item
  (\emph{Winning Region})
Given an LTL game $\myGame$, compute the winning region for
Player~$C$. Wherever possible, also compute a finite-memory winning
strategy for $C$ from this region.
\item 
  (\emph{Realizability})
Given an LTL game $\myGame$ and an initial region in the form of a
constraint $\init$ over the variables $V$ of the game, decide whether
Player~$C$ wins from every state in $\init$. If possible, compute a
finite-memory winning strategy for $C$ from $\init$.
\end{enumerate}

It is easy to see that these problems are undecidable in general (for
example by a reduction from the control-state reachability problem for
2-Counter Machines).
Hence the procedures we give in subsequent sections may not always be
terminating ones.
In the sequel we focus on the problem of computing winning regions,
since we can check realizability by checking if the given initial
region is contained in the winning region.


\section{\gensysltl\ Approach}

\begin{algorithm}
\SetKwComment{Comment}{/\!/ }{ }
\SetKwInOut{Input}{Input}
\SetKwInOut{Output}{Output}
\SetKwRepeat{Do}{do}{while}
\Input{ LTL game $\myGame = (V, D, \cont, \envi, \psi)$ }
\Output{$\winreg{C}{\myGame}$ or an approximation
  of it, and a strategy $\sigma$
  for Player~$C$ from this region.}
\If { $\mathcal{G}$ is simple } {
  Compute $\winreg{C}{\myGame}$ (i.e.\@ winning reg for $C$ in
  $\myGame$). \\
  Compute winning strategy $\sigma$. \\
  \Return $\winreg{C}{\myGame}$, $\sigma$.
}

$\A_{\psi}$ := $\ltltba(\psi)$. \\
$\A_{\neg\psi}$ := $\ltltba(\neg\psi)$. \\

\If { $\A_{\psi}$ is deterministic }{ Construct simple B\"uchi product game
  $\mathcal{H} = \myGame \myprod \A_\psi$. \\
  Compute $\winreg{C}{\mathcal{H}}$. \\
  Extract $\winreg{C}{\myGame}$, winning strategy
  $\sigma$ for $C$ in $\myGame$. \\
\Return $\winreg{C}{\myGame}$, $\sigma$.    
}

\If { $\A_{\neg\psi}$ is deterministic }{
  Construct simple Co-B\"uchi product game $\mathcal{H} = \myGame \myprod
  \A_{\neg\psi}$.
  Compute $\winreg{C}{\mathcal{H}}$. \\
  Extract $\winreg{C}{\myGame}$, winning strategy $\sigma$ for $C$. \\
\Return $\winreg{C}{\myGame}$, $\sigma$.    
}

\Comment{Both $\A_\psi$ and $\A_{\neg\psi}$ are non-det}
$k := 0$; $W_U := \false$; $W_O := \true$; \\
\While { $W_U \neq W_O$} {
  Construct on-the-fly two $k$-safety product automatons involving $\myGame$ with $\apsi$ and $\anpsi$, respectively, and from these, extract an under-approximation $W_U$ of $\winreg{C}{\myGame}$ and an over-approximation $W_O$ of $\winreg{C}{\myGame}$, respectively. From $W_U$ extract a winning strategy $\sigma$ for Player~$C$. 
  
  


   
  
  
  $k$ = $k$ + 1.
}
\Return $W_U, W_O, \sigma$;
\caption{\gensysltl\ overview}
\label{proc:gensys-ltl}
\end{algorithm}

\begin{figure}
  \centering
  \includegraphics[scale=0.45]{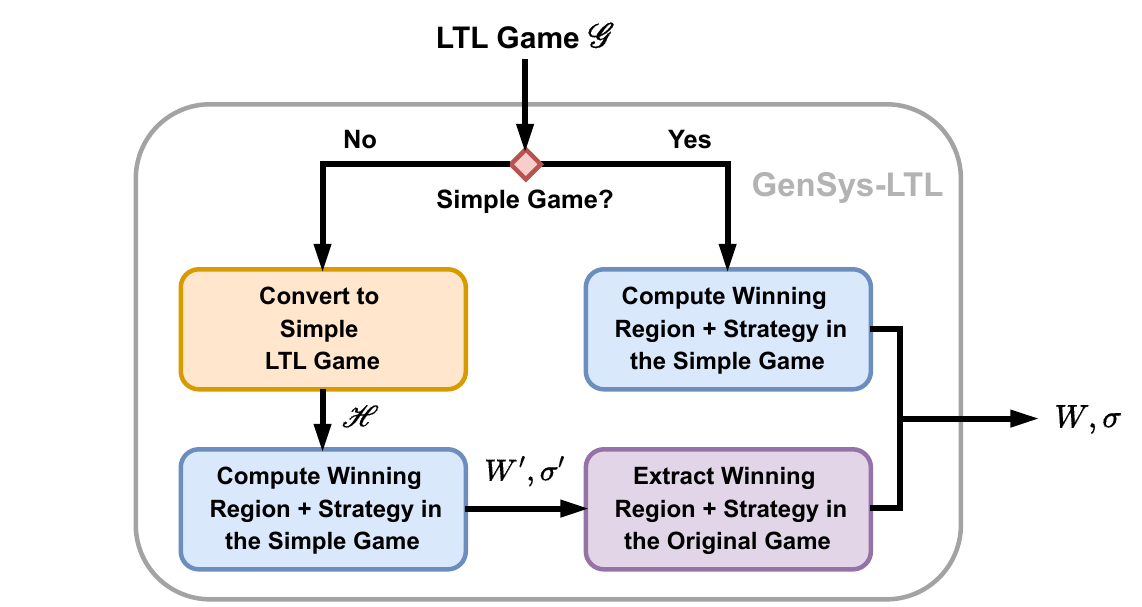}
   \caption{Schematic overview of \gensysltl}
  \label{fig:gensysltl-schematic}
\end{figure}

Our approach consists of a bouquet of techniques. This is motivated by our
 objective to
handle  each type of LTL formula with an efficient technique suited to
that type. Algorithm~\ref{proc:gensys-ltl} is the ``main'' program or
driver of our approach. Fig.~\ref{fig:gensysltl-schematic} also
summarizes the approach.

Algorithm~\ref{proc:gensys-ltl} takes as input an
LTL game $\mathcal{G}$.
Lines~1--4 of the algorithm tackle the scenario when the given game
$\myGame$ is \emph{simple}. These are games where the formula
$\psi$ is of one of restricted forms $\Gl (X)$ (safety), $\F (X)$
(reachability), $\Gl \F (X)$ (B\"uchi), or
$\F \Gl (X)$ (Co-B\"uchi), where $X$ is a constraint over the game variables
$V$. For these cases, we propose fixpoint procedures
that directly work on the state-space of
the given game $\myGame$, and that use SMT formulae to encode sets of
states. Sec.~\ref{sec:simple-games} describes these procedures in detail.
Due to the infiniteness of the state-space, these fixpoint
computations are not guaranteed to terminate. When they do terminate, they
are guaranteed to compute the precise winning region $\winreg{C}{\myGame}$,
and, in the case of FND games, to extract a memoryless strategy
automaton for these regions.

If the given formula $\psi$ is not simple, we convert the formula as well
as its negation, in Lines~5--6, into B\"uchi automata using a standard
procedure~\cite{WolperVS83}. If either of these two automata are
\emph{deterministic} (see Sec.~\ref{sec:ltl}), we 
construct a product of the game $\myGame$ with the automaton, such that
this product-game $\mathcal{H}$ is a simple LTL game. We then compute the
winning region on this product using the fixpoint computations mentioned
above. If the fixpoint computation terminates, we  extract a winning
region for the
original game $\myGame$, and a strategy $\sigma$. These steps
are outlined in Lines~7--15 of Algorithm~\ref{proc:gensys-ltl}, and details
are presented in Sec.~\ref{sec:deterministic-games}.

The hardest scenario is when both the automata are non-deterministic.  For
this scenario, we propose an on-the-fly determinization and winning-region
extraction approach. These steps are outlined in Lines~16-19 of
Algorithm~\ref{proc:gensys-ltl}.
We
present the details  in Section~\ref{sec:otf}. 

We state the following claim which we will substantiate in subsequent
sections:

\begin{theorem}
  \label{thm:gensysltl-approach}
  \itshape{
    Whenever Algorithm~\ref{proc:gensys-ltl} terminates, it outputs the
exact winning region for Player~$C$ when $\myGame$ is either simple or
deterministic; in other cases it outputs a sound under- and
over-approximation of the winning region for Player~$C$ in $\myGame$.
Additionally, in the case when $\myGame$ is FND,
upon termination Algorithm~\ref{proc:gensys-ltl} outputs a strategy
automaton representing a winning strategy for Player~$C$ from this
region.} \hfill \qed
\end{theorem}

\section{Simple LTL Games}
\label{sec:simple-games}

Our approach reduces logical LTL games to ``simple'' LTL games in which
the winning condition is \emph{internal} to the game.
In this section we describe this subclass of LTL games and the basic
fixpoint algorithms to solve them.

A 
\emph{simple} LTL game is an LTL game 
\(
\myGame = (V, D, \cont, \envi, \psi),
\)
in which $\psi$ is an $\ltl(V)$ formula of the form $\Gl(X)$, $\F(X)$,
$\Gl\F(X)$, or 
$\F\Gl(X)$, where $X$ is a constraint on $V$.
We refer to games with such specifications as \emph{safety},
\emph{reachability}, \emph{B\"uchi}, and \emph{co-B\"uchi} games,
respectively.

We can compute (with the possibility of non-termination) the winning
regions $\winreg{C}{\myGame}$ for each of these four types of games,
and a strategy automaton representing a memoryless winning strategy
for Player~$C$, in the special case of 
FND 
games, by extending the classical
algorithms for the finite-state versions of these games
(see \cite{Thomas1995,MalerPS95}).



The algorithms we describe will make use of the following formulas
representing sets of ``controllable predecessors'' in the context of
different types of games.
Here $Y(V)$ is a constraint representing a set of game states.
\begin{itemize}
\item The set of
  controllable predecessors w.r.t.\@ the set of 
  states $Y$, for a safety specification $\Gl(X)$ (namely 
  states from which Player~$C$ has
  a \emph{safe} move from which all environment moves result in a $Y$-state):
$$
\begin{array}{lcl}
\wpr_S^X(Y)& \equiv & \exists V' (\cont(V,V') \ \wedge X(V') \ \wedge\\
&& \ \forall V'' (
\envi(V', V'') \implies  Y(V''))).
\end{array}
$$

\item The set of controllable predecessors w.r.t.\@ $Y$, for a
  reachability specification $\F(X)$ (namely states from which either
  $C$ has a move that either gets into $X$, or from which all
  environment moves get into $Y$):
 $$
 \begin{array}{lcl}
 \wpr_R^X(Y)& \equiv & \exists V' (\cont(V,V') \ \wedge (X(V') \ \vee\\
 && \ \forall V'' (
 \envi(V', V'') \implies  Y(V'')))).
 \end{array}
 $$
 
 \item The set of predecessors w.r.t.\@ $Y$ for Player $C$ (namely states from which 
  $C$ has a move that results in a $Y$-state):
 $$
 \begin{array}{lcl}
 \wpr_C(Y)& \equiv & \exists V' (\cont(V,V') \ \wedge  Y(V')).
 \end{array}
 $$
 
 \item The set of predecessors w.r.t.\@ $Y$ for Player $E$ (namely states from which 
  all moves of $E$ result in a $Y$-state):
 $$
 \begin{array}{lcl}
 \wpr_E(Y)& \equiv &\forall V' ( \envi(V, V') \implies  Y(V')).
 \end{array}
 $$
 
%

\end{itemize}

\begin{algorithm}
\SetKwComment{Comment}{/* }{ */}
\SetKwInOut{Input}{Input}
\SetKwInOut{Output}{Output}
\SetKwRepeat{Do}{do}{while}
\Input{ Safety game $\myGame = (V, D, \cont, \envi, \Gl(X))$ }
\Output{ $\winreg{C}{\myGame}$, strategy $\sigma$}

$W$ := $X$;

\Do{$\neg(W_{\mathit{old}} \Rightarrow W)$}{ 

  $W_{\mathit{old}}$ := $W$; 

  $W$ := $\qelim(\wpr_S^X(W) \wedge X)$;

}

$\sigma$ := $\mathit{ExtractStrategy}_\Gl(W)$;

\Return $W$, $\sigma$;
\caption{ComputeWR-Safety}
\label{proc:gensys-safety}
\end{algorithm}

\begin{algorithm}
\SetKwComment{Comment}{/* }{ */}
\SetKwInOut{Input}{Input}
\SetKwInOut{Output}{Output}
\SetKwRepeat{Do}{do}{while}
\Input{ Reachability game $\myGame = (V, D, \cont, \envi, \F (X))$ }
\Output{ $\winreg{C}{\myGame}$, strategy $\sigma$}

$W$ := $X$;

$C$ := $[W]$;

\Do{$\neg(W \Rightarrow W_{\mathit{old}})$}{ 

  $W_{\mathit{old}}$ := $W$; 

  $W$ := $\qelim(\wpr_R^X(W) \vee X)$;

  $C.\mathit{append}(W \wedge \neg W_{\mathit{old}})$;
}

$\sigma$ := $\mathit{ExtractStrategy}_\F(C)$;

\Return $W$, $\sigma$;
\caption{ComputeWR-Reachability}
\label{proc:gensys-reachability}
\end{algorithm}

\begin{algorithm}
\SetKwComment{Comment}{/* }{ */}
\SetKwInOut{Input}{Input}
\SetKwInOut{Output}{Output}
\SetKwRepeat{Do}{do}{while}
\Input{ B\"uchi game $\myGame = (V, D, \cont, \envi, \Gl \F (X))$ }
\Output{ $\winreg{C}{\myGame}$, strategy $\sigma$}

$W$ := $W_E$ := $True$;

\Do{$\neg (W_{E_{\mathit{old}}} \Rightarrow W_E \ \wedge \ W_{\mathit{old}} \Rightarrow W)$}{ 

$W_{E_{\mathit{old}}}, W_{\mathit{old}}$ := $W_E, W$;


$W$ := $\qelim(\wpr_C(W_E) \wedge X)$;

$W_E$ := $\qelim(\wpr_E(W) \wedge X)$;

$H$ := $H_E$ := $False$;

$C$ := $[H]$;

    \Do{$\neg (H_E \Rightarrow H_{E_{\mathit{old}}} \ \wedge \ H \Rightarrow H_{\mathit{old}})$}{ 
    
    $H_{E_{\mathit{old}}}, H_{\mathit{old}}$ := $H_E, H$;


	$H$ := $\qelim(\wpr_C(H_E) \vee W)$;

	$H_E$ := $\qelim(\wpr_E(H) \vee W_E)$;
	
	 $C.\mathit{append}(H \wedge \neg H_{\mathit{old}})$;
	 
    }

$W_E,  W$ := $H_E,  H$;


}

$\sigma$ := $ExtractStrategy_{GF}(W , C)$;

\Return $W,  \sigma$;

\caption{ComputeWR - B\"uchi}
\label{proc:gensys-buchi}
\end{algorithm}

%
%
%
%
%
%
%
%
%
%
%
%
%
%
%
%

\begin{algorithm}
\SetKwComment{Comment}{/* }{ */}
\SetKwInOut{Input}{Input}
\SetKwInOut{Output}{Output}
\SetKwRepeat{Do}{do}{while}
\Input{ Co-B\"uchi game $\myGame = (V, D, \cont, \envi, \F \Gl(X))$ }
\Output{ $\winreg{C}{\myGame}$, strategy $\sigma$}

$W$ := $W_E$ := $False$;

$C$ := $[W]$;

\Do{$\neg (W_E \Rightarrow W_{E_{\mathit{old}}} \ \wedge \ W \Rightarrow W_{\mathit{old}})$}{ 

$W_{E_{\mathit{old}}}, W_{\mathit{old}}$ := $W_E, W$;


$W$ := $\qelim(\wpr_C(W_E) \vee X)$;

$W_E$ := $\qelim(\wpr_E(W) \vee X)$;

$H$ := $H_E$ := $True$;

    \Do{$\neg (H_{E_{\mathit{old}}} \Rightarrow H_E \ \wedge \ H_{\mathit{old}} \Rightarrow H)$}{ 
    
    $H_{E_{\mathit{old}}}, H_{\mathit{old}}$ := $H_E, H$;


	$H$ := $\qelim(\wpr_C(H_E) \wedge W)$;

	$H_E$ := $\qelim(\wpr_E(H) \wedge W_E)$;
	 
    }

$W_E,  W$ := $H_E, H$;


$C.\mathit{append}(W \wedge \neg W_{\mathit{old}})$;
}

$\sigma$ := $ExtractStrategy_{GF}(W , C)$;

\Return $W,  \sigma$;
\caption{ComputeWR - Co-B\"uchi}
\label{proc:gensys-cobuchi}
\end{algorithm}

Algorithm~\ref{proc:gensys-safety} (ComputeWR-Safety) takes a safety
game as input, and
iteratively computes the safe controllable predecessors,
starting with the given safe set $X$, until it reaches a fixpoint
($W_\mathit{old} \implies W$).
Here we use a quantifier elimination procedure $\qelim$ which takes a
logical formula with quantifiers (like $\wpr_S^X(W) \myAnd X$) and
returns an equivalent quantifier-free formula.  For example, 
$\qelim(\exists y (y \leq x \myAnd x + y \leq 1 \myAnd 0 \leq y))$
returns $0 \leq x \myAnd x \leq 1$.
Upon termination the algorithm returns the fixpoint $W$.

When the input game is FND (with $\cont = \cont_1 \myOr \cdots
\myOr \cont_k$),
the call to
$\mathit{ExtractStrategy}_G(W)$ does the following.
Let
\[
\begin{array}{lll}
  U_i = & W \wedge \qelim(\exists V' (& \cont_i(V,V') \ \wedge W(V') \ \wedge
  \\
        &                    & \forall V'' (\envi(V', V'') \implies  W(V'')))).
\end{array}
\]
Then the memoryless strategy $\sigma$ extracted simply offers the move
$\cont_i$ whenever Player~$C$ is in region $U_i$.
The corresponding strategy automaton essentially maintains a
controller state for each constraint $U_i$, labelled by the move $\cont_i$.  For the strategy extraction in the rest of this section,  we assume that the input game is FND.

Similarly,  Algorithm~\ref{proc:gensys-reachability} (ComputeWR-Reachability) takes a reachability
game as input, and
iteratively computes the reachable controllable predecessors,
starting with the given safe set $X$, until it reaches a fixpoint
($W \implies W_\mathit{old}$).   


To compute the memoryless strategy for reachability,  we compute $C$ that ensures that each move made by the controller from a given state ensures that it moves one step closer to $X$.  

Let the reachability controllable predecessor for move $\cont_i$ be:
 $$
 \begin{array}{lcl}
 \wpr_{R_i}^X(Y)& \equiv & \exists V' (\cont_i(V,V') \ \wedge (X(V') \ \vee\\
 && \ \forall V'' (
 \envi(V', V'') \implies  Y(V'')))).
 \end{array}
 $$

Then $\mathit{ExtractStrategy}_\F(C)$ does the following:  
\[
\begin{array}{lll}
  U_i = & \bigvee_{j = 0}^{|C| -2} \qelim(\wpr_{R_i}^{X_j}(X_j)) \wedge C_{j+1}
\end{array}
\]
where $X_j = C_j \vee C_{j-1} \vee C_{j-2} \vee \cdots \vee C_0$.

Thus,  $U_i$ is the set of states exclusively in $W_{j+1}$ (which we denote by $C_{j-1}$ which is constructed in Algorithm \ref{proc:gensys-reachability}) from where Player $C$ has a strategy to reach $X$ by first ensuring a move to $W_j$, thereby ensuring moving one step closer to $X$.  Then the memoryless strategy $\sigma$ extracted offers the move $\cont_i$ whenever Player~$C$ is in the region $U_i$.
The corresponding strategy automaton essentially maintains a
controller state for each constraint $U_i$, labelled by the move $\cont_i$.


Algorithm~\ref{proc:gensys-buchi} (ComputeWR- B\"uchi) takes a B\"uchi game as an input,  and computes a winning region from where Player $C$ has a strategy to visit $X$ infinitely often.  In this algorithm,  we require two levels of nesting to compute the winning region.  Using two-step controllable predecessors
 (such as $\wpr_S^X$,  and $\wpr_R^X$),  that reason about two moves at a time causes unsoundness,  if used directly.  Using $\wpr_S^X$ in the nested Buchi algorithm causes an underapproximation of the winning region as it is not necessary that the intermediate environment states be safe. 
  Similarly,  using $\wpr_R^X$ is too weak as the intermediate states of the environment are not reasoned with correctly.  It assumes that a finite play reaching an intermediate environment state in $X$ satisfies the property,  which is not true for an infinite B\"uchi play.  
Thus,  we use one step controllable predecessors $\wpr_C$ and $\wpr_E$ (for controller and environment respectively) that reason about the game play one move at a time in the style of \cite{Thomas1995}.  The strategy is also extracted similarly.

As a dual of Algorithm~\ref{proc:gensys-buchi},  Algorithm~\ref{proc:gensys-cobuchi} (ComputeWR- Co-B\"uchi),  takes a co-b\"uchi game as an input,  and computes a winning region from where Player $C$ has a strategy to eventually visit $X$ always.

We can now state (see App.~\ref{sec:appendix-simple-games-proof} for proof):


\begin{theorem}
  \label{thm:simple-games}
\itshape{
Whenever
Algorithms~\ref{proc:gensys-safety}, \ref{proc:gensys-reachability},
\ref{proc:gensys-buchi}, and \ref{proc:gensys-cobuchi}
terminate, they compute the exact winning region for Player~$C$ in
safety, reachability, B\"uchi, and co-B\"uchi games, respectively.
For FND games, upon termination, they also output a winning
strategy automaton for Player~$C$ for this region.
Furthermore, for safety games this strategy is maximally permissive.}
\hfill \qed
\end{theorem}

\section{Deterministic LTL Games}
\label{sec:deterministic-games}

In this section we discuss how to solve a game with an LTL
condition $\psi$ which is not simple, but is nevertheless
\emph{deterministic} in that
either $\A_\psi$ or $\A_{\neg\psi}$ is deterministic.
We begin with the case when $\A_\psi$ is deterministic.

Let $\myGame = (V, D, \cont, \envi, \psi)$ be an LTL game, and let
$\A_\psi = (Q, \{q_0\}, \T, F)$ be a deterministic and complete
B\"uchi automaton for $\psi$ over the set of variables $V$.
We define the \emph{product game} corresponding to $\myGame$ and $\A_\psi$ to be
\[
\myGame \myprod \A_{\psi} = (V \union\{q\}, D, \cont', \envi', \psi')
\ \ \ \textrm{where}
\]
\begin{itemize}
\item $q$ is a new variable representing the state of the automaton such that $D(q) =\{1, 2, \cdots |Q|\}$ 
\item $\cont' = \cont \myAnd \bigvee_{(p,\delta,p') \in \T} (q = p
   \myAnd \delta \myAnd q' = p')$.
\item $\envi' = \envi \myAnd \bigvee_{(p,\delta,p') \in \T} (q = p
   \myAnd \delta \myAnd q' = p')$.
\item $\psi' = \Gl\F(\bigvee_{p \in F} q = p)$.
\end{itemize}

Similarly,  for the case when $\A_{\neg\psi}$ is a deterministic and complete B\"uchi automaton for $\neg\psi$,    we define the \emph{product game} corresponding to $\myGame$ and $\A_{\neg\psi}$ to be
\[
\myGame \myprod \A_{\neg\psi} = (V \union\{q\}, D, \cont', \envi', \psi')
\ \ \ \textrm{where}
\]

\begin{itemize}
\item $\psi' = \F\Gl(\bigvee_{p \notin F} q = p)$.
\end{itemize}

The definitions of $\cont'$ and $\envi'$ remain the same as that of the product game $\myGame \myprod \A_{\psi}$.  For the product game $\myGame \myprod \A_{\neg\psi}$,  in order to satisfy the specification $\psi$, we need to visit the final states of $\A_{\neg\psi}$ finitely often. This is equivalent to visiting the non-final states eventually always,  as the definition of $\psi'$ states. 
We note that if $\myGame$ is finitely non-deterministic, so is $\myGame \myprod \A_{\psi}$ and
$\myGame \myprod \A_{\neg\psi}$.

\begin{theorem}
\label{thm:deterministic-games}
\itshape{
Let $\myGame$, with $\A_\psi$ (resp. $\A_{\neg\psi}$) deterministic, be as above.
Let $W'$ be the winning region for Player~$C$ in $\myGame \myprod \A_\psi$ (resp.  $\myGame \myprod \A_{\neg\psi}$).
Then the winning region for Player~$C$ in $\myGame$ is
$W = \{ s \ | \ (s,q_0) \in W' \}$.
Furthermore, when $\myGame$ is finitely non-deterministic, given a
finitely-representable memoryless strategy for $C$ in $\myGame \myprod
\A_{\psi}$ (resp.  $\myGame \myprod \A_{\neg\psi}$), we
can construct a 
finitely-representable finite-memory strategy for $C$ in $\myGame$.}
\end{theorem}
\begin{IEEEproof}
See Appendix~\ref{sec:appendix-deterministic-games-proof}.
\end{IEEEproof}

\section{On The Fly Determinization Approach}
\label{sec:otf}

When the automata $\apsi$ and $\anpsi$ are non-deterministic, the product game  $\hpsi$ of the given game $\myGame$ with $\apsi$ and the product game $\hnpsi$ of $\myGame$ with $\anpsi$ both will be non-deterministic. It has been recognized in the literature~\cite{jobstmann2018graph} that non-deterministic automata need  to be determinized to enable a precise winning region to be inferred.


\subsection{Overview of determinization}
\label{ssec:overview-otf}

\begin{figure}
  \centering
  \includegraphics[scale=0.35]{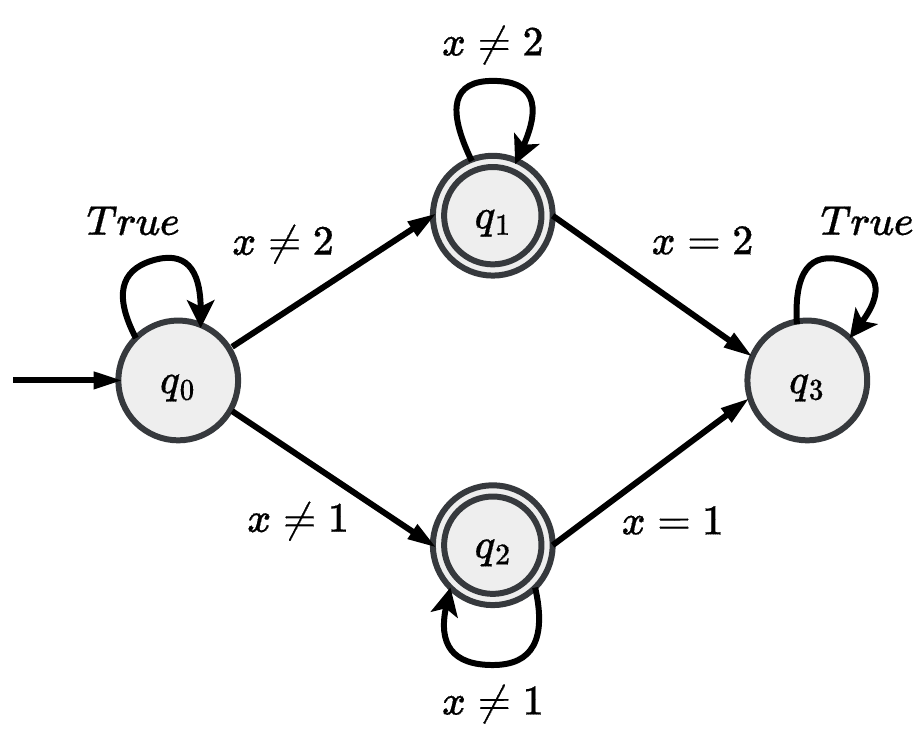}
  \caption{Universal Co-B\"uchi automaton $\anpsi$ for the specification $\psi$ := $ \Gl (\F (x = 1) \wedge \F (x = 2))$}
  \label{fig:ucw1-example}
\end{figure}

\begin{figure*}
  \centering
  \includegraphics[scale=0.45]{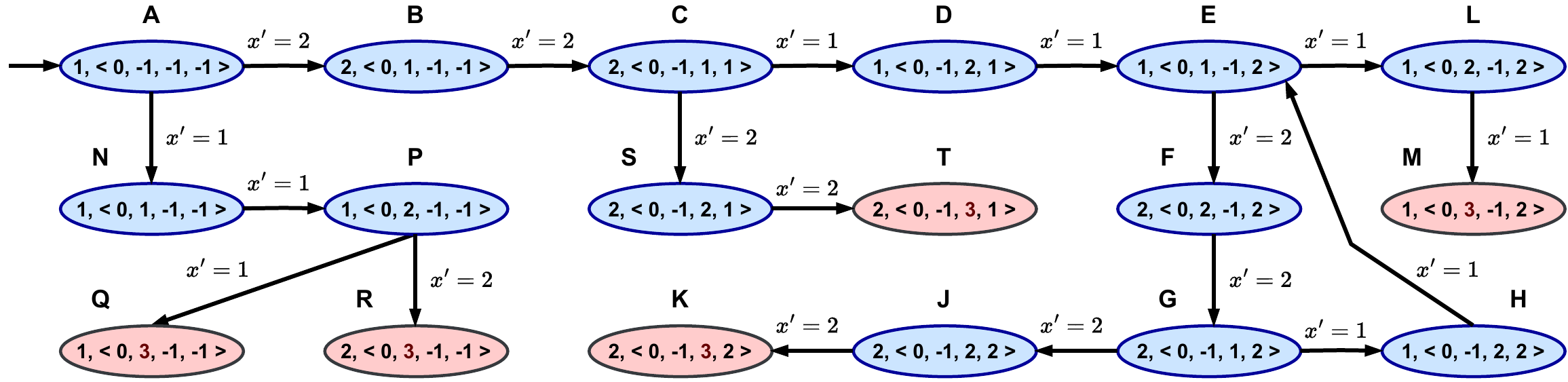}
  \caption{Determinized 2-safety game for $\anpsi$ where $\psi$ :=  $\Gl (\F (x = 1) \wedge \F (x = 2))$}
  \label{fig:ksafety-example}
\end{figure*}

We adopt the basic idea of $k$-safety determinization from the
\emph{Acacia} approach~\cite{filiot2011antichains} for finite games
and extend it to the setting of infinite games. We introduce our
determinized product game construction intuitively below, and later
formally in Sec.~\ref{ssec:formal-otf}. The underlying game graph
$\myGame$ we use here for illustration is based on the Elevator game
in Example~\ref{example:elevator}.
We simplify the example to admit just two controller moves, namely, $x' = 1$ and $x' = 2$, while the environment does not change the floor $x$ in its moves. The given LTL property $\psi$ is $\Gl (\F (x = 1) \wedge \F (x = 2))$. Fig.~\ref{fig:ucw1-example} depicts the Universal Co-B\"uchi automaton $\anpsi$ for this property, which happens to be non-deterministic.

The approach takes as parameter an  integer $k \geq 0$, and generates
a determinized version of the product of the game $\myGame$ and the
automaton $\anpsi$. A portion of the (infinite) determinized product
for our example under consideration is depicted in
Fig.~\ref{fig:ksafety-example}, for $k=2$. Each state of the
determinized product is a pair of the form $(s, v)$, where $s$ a state
of the underlying game $\myGame$ (i.e., a value of $x$ in the
example), and $v$ is a vector of counts (the vectors are depicted
within angle brackets).  Each vector intuitively represents the subset
of automaton states that the game could be in currently,  with $v(i) >
-1$ indicating that the automaton state $q_i \in Q$ belongs to the
subset. If $v(i) > -1$, the value $v(i)$ further indicates the  count
of the maximum number of final states that can be visited along plays
in the underlying game graph that reach automaton state $q_i$ and that
correspond to plays of the product graph that reach the current state
$(s, v)$. The moves of the two players in the product graph are
alternating.  For conciseness,  we avoid showing the environment states in square., which do not make any updates to the game state. The initial states of the product graph are the ones
whose vector component is  $c_0 = \langle 0,-1,-1,-1 \rangle$, which represents the \emph{initial} subset $\{q_0\}$. One of the initial states of the product graph is depicted in Fig.~\ref{fig:ksafety-example} (there are an infinite number of them, corresponding to all possible values of the game variable $x$).

We pick state~$E$ in Fig.~\ref{fig:ksafety-example} to illustrate the subset construction.  $q_2$ is not present in~$E$ because from none of the automaton states that are present in the subset in product state~$D$ (i.e.,  $q_0, q_2$ or $q_3$), transitions to $q_2$ are possible as per the automaton in Fig.~\ref{fig:ucw1-example},  when the value of $x$ is 1 (as $x$ has value~1 in product state~$D$). And $q_3$  has a count of~2 in $E$ because $q_2$ had count of~2 in state~$D$ and a $q_2$ to $q_3$ transition is possible when $x$ has value~1 as per Fig.~\ref{fig:ucw1-example}.


The product game shown in Fig.~\ref{fig:ksafety-example} can be seen to deterministic. This means that if a product state $(s, v)$ has two successors $(s_1, v_1)$ and $(s_2, v_2)$, then $s_1 \neq s_2$.
\emph{Safe} states of the product graph are ones where no element of the vector exceeds $k$.  Successor states of unsafe states will themselves be unsafe. In the figure unsafe states are indicated in red color (and have entries greater than 2 in the vectors).

A unique product game graph exists as per the construction mentioned
above for a given value of $k$. This product game graph is said to be
\emph{winning} for Player~$C$ if satisfies the following conditions:
(I) At least one of the safe product states has $c_0 = \langle 0,-1,\ldots,-1\rangle$ as its vector,  (II) For every safe product state from which the controller moves, at least one of the successors is a safe state, and (III) for every safe product state from which the environment moves, none of the successors are unsafe states. Otherwise, higher values of $k$ will need to be tried, as indicated in the loop in Lines~16-19 in Algorithm~\ref{proc:gensys-ltl}. The product game graph in Fig.~\ref{fig:ksafety-example} happens to be winning. 

If a product game graph is winning,  then the game state components
$s$ of the product states of the form $(s, c_0)$ in the product graph,
where $c_0 = \langle 0, -1, \ldots, -1\rangle$,  constitute, in
general, an under-approximation of the winning region
$\winreg{C}{\myGame}$. The under-approximation in general increases in
size as the value of $k$ increases. Note, in the loop we also compute
a strategy for Player~$E$ by constructing a determinized product using
$\apsi$. Using this  it can be detected when the current value of $k$
yields the precise region $\winreg{C}{\myGame}$. (If the underlying
game 
is finite, such a $k$ is guaranteed to exist.)




\subsection{Formal presentation of deterministic product construction}
\label{ssec:formal-otf}

We present here our SMT-based fixpoint computation for computing the product game graph of the kind introduced above, for a given bound $k$. We use formulas to represent (finite or infinite) portions of product graphs symbolically. The free variables in any formula are underlying game variables $V$ and a vector-typed variable $c$.  The solution to a formula is a (finite or infinite) set of  states of a product graph.

$\aut(P, V, Q)$ is a given formula that encodes the logical transition relation $\T$ of the B\"uchi automaton $\anpsi$.
A triple $(q, s, q')$ is   a solution to $\aut(P, V, Q)$ iff $(q, s, q') \in \Delta_{\T}$. For instance, for the automaton in Fig.~\ref{fig:ucw1-example}, $\aut(P, V, Q)$ would be $(P = q_0 \wedge Q = q_0) \vee (P = q_0 \wedge Q = q_1 \wedge x \neq 2) \vee \cdots$. $\fin(P)$  is a given formula that evaluates to~1 if $P$ is a final state in the automaton $\anpsi$ and otherwise evaluates to~0. 


We define a formula $\succc_k(c,V,c')$ as follows:
\begin{align*}
  \forall q. \, c'(q) & = \mathit{max}\{\mathit{min}(c(p) + \fin(q), k+1) \, | \\
   & \hspace*{1.5cm} p \in Q, \aut(p,V,q), c(p) \geq 0\},  \\
   & \hspace*{0.5cm} \mathrm{if\ } \exists p \mathrm{\ such\ that\ }  \aut(p,V,q) \wedge c(p) \geq 0\\ 
  &= -1, \mbox{ otherwise.}
\end{align*}

Intuitively, a triple $(v, s, v')$ is a solution to $\succc_k(c,V,c')$ iff the product state $(s', v')$ is a successor of the product state $(s, v)$ for some $s'$.

Our approach is to use an iterative shrinking fixpoint computation to compute the \emph{greatest fixpoint} (GFP) $W$ of the function $\wpr$ defined below.
\begin{align*}
\wpr_k(X) \ \equiv \ &  G(V,c) \, \wedge \\
& \exists V',c'.\ (\cont(V,V')  \wedge \succc_k(c, V, c') \wedge \,  G(V',c')  \\
& \ \ \wedge  \forall V'',c''. \  (
(Env(V', V'')  \wedge  \succc_k(c', V, c''))  \\
& \hspace*{2cm} \implies  X(V'',c''))).
\end{align*}

The argument of and the return value from the function above are both
formulas in $V,c$, representing sets of product graph states.
$G(V,c)$ represents safe product states, and checks that all elements
of $c$ are $\geq$ -1  and $\leq k$. The fixpoint computation is not
guaranteed to terminate due to the infiniteness in the underlying game
graph $\myGame$. If it does terminate, the formula $W$, after
replacing  the free variable $c$  with the initial vector $c_0 =
\langle 0,-1,\ldots,-1\rangle$, is returned.  This formula will have solutions iff the value of $k$ considered was sufficient to identify a non-empty under-approximation of $\winreg{C}{\myGame}$. The formula's solution is guaranteed to represent the \emph{maximal} winning product graph that exists (and hence the maximal subset of $\winreg{C}{\myGame}$) for the given value of $k$.

If $W$ has solutions, we infer a strategy $\sigma$ for Player~$C$ as follows. 
The following utility function $\sigma_\mathit{prod}$ returns a formula in free variables $V$, whose solutions are the next game states to transition to when at a product state $(s_1,c_1)$ in order to ensure a winning play. 
$$
\sigma_{\mathit{prod}}(s_1,c_1) \ = \ \exists c_2. \, \cont(s_1,V) \wedge \succc_k(c_1,s_1,c_2) \wedge W(V,c_2)
$$

We introduce a utility function $\destp$, whose argument is a play in the underlying game $\myGame$, and that returns the product state in the determinized product graph reached by the play.
\begin{align*}
  \destp(s) & =  (s, c_0) \\
  \destp(w \cdot s) & =  (s, c), \mbox{ such that } \\
  & \hspace*{.5cm} (\destp(w) = (s', c') \wedge \succc_k(c', s', c)
\end{align*}

Finally, the strategy in terms of the underlying game $\myGame$ is defined as follows (where $w$ is a play in the underlying game):
$$
\sigma(w) \ = \ \sigma_\mathit{prod}(\destp(w)).
$$

\section{Implementation}

We implement all fixpoint approaches in our protoype tool \gensysltl\, which extends our earlier tool GenSys \cite{gensys} to support general LTL specifications.  \gensysltl\ is implemented using Python and uses the Z3 theorem prover \cite{z3} from Microsoft Research as the constraint solver under the hood. \gensysltl\ uses Z3 to eliminate quantifiers from formulas resulting from the fixpoint iterations and check satisfiability.  In all fixpoint approaches mentioned in this paper,  large formulas are generated in every iteration,  containing nested quantifiers. This formula blowup can quickly cause a bottleneck affecting scalability. The reason is that Z3 chokes over large formulas involving nested quantifiers. Thus,  it is necessary to eliminate quantifiers at every step. We use quantifier elimination tactics inherent in Z3 to solve this issue.  We use variations of \cite{qe2} and simplification tactics in parallel,  to achieve efficient quantifier elimination. 


To convert a given LTL formula into an equivalent Buchi automaton, we use the Spot library \cite{spot} which efficiently returns a complete and state based accepting automaton for a given LTL specification.  We also constrain Spot to return a deterministic Buchi automaton whenever possible,  and then choose our approach appropriately.  However,  in this prototype version of \gensysltl\ ,  this encoding is done manually.  \gensysltl\ is available as an open source tool on GitHub\footnote{\url{https://github.com/stanlysamuel/gensys/tree/gensys-ltl}}.
\section{Evaluation}

To evaluate \gensysltl\, we collect  from the literature a corpus of benchmarks (and corresponding temporal specifications) that deal with the synthesis of strategies for two-player infinite-state logical LTL games.  The first set of benchmarks were used in the evaluation of the \emph{ConSynth}~\cite{consynth} approach. These target program repair applications,  program synchronization and synthesis scenarios for single and multi-threaded programs,  and variations of the  Cinderella-Stepmother game \cite{cinderella1, cinderella2},  which is
considered to be a challenging program for automated synthesis tasks.  The second set of benchmarks were used to evaluate the \emph{Raboniel}~\cite{raboniel} approach, which contains elevator, sorting,  and cyber-physical examples, and specification complexity ranging from simple LTL games to ones that need products with B\"uchi automata. 
The third set of benchmarks are from \emph{DTSynth} approach evaluation~\cite{dtsynth}, and involve safety properties on  robot motion planning  over an infinite grid. 
 
We compare our tool {\gensysltl} against two comparable tools from the literature: ConSynth~\cite{consynth}  and Raboniel~\cite{raboniel}. We do not compare against tools such as DTSynth~\cite{dtsynth} that only handle safety (not general LTL) specifications. We executed {\gensysltl} and Raboniel on our benchmarks on a desktop computer  having six Intel i7-8700 cores at 3.20GHz each and  64 GB RAM. We were able to obtain a binary for  ConSynth from other authors~\cite{dtsynth}, but were unable to  run it due to incompatibilities with numerous versions of Z3 that we tried with it. Hence, for the benchmarks in our suite that previous papers~\cite{dtsynth,consynth} had evaluated ConSynth on, we directly picked up results from those papers. There is another comparable synthesis tool we are aware of, \emph{Temos}~\cite{temos}. We were unable to install this tool successfully from their code available on their artifact and from their GitHub repository, due to numerous dependencies that we could not successfully resolve despite much effort.

Table \ref{table: results} shows the experimental results of all our approaches in comparison with ConSynth and Raboniel,  with a timeout of 15 minutes per  benchmark. The first column depicts the name of the benchmark: each benchmark includes a logical game specification and a temporal property winning condition. Column~\textbf{Type} indicates whether the game variables in the underlying game $\myGame$ are reals or integers. Column~\textbf{P} indicates the player ($C$ or $E$) for which we are synthesizing a winning region. Column~\textbf{S} indicates whether the given benchmark falls in the Simple LTL category (G, F, FG, GF), or whether it needs an automaton to be constructed from the LTL property (Gen).   $|$V$|$ is the number of game variables. Letting $\psi$ denote the given temporal property, Column~\textbf{DB?} indicates whether the automaton $\apsi$ is deterministic, while  Column~\textbf{DCB?} indicates whether the automaton $\anpsi$ is deterministic. In both these columns, the numbers within brackets indicate the number of automaton states. 

The remainder of this section summarizes our results for the two main problems we address in this paper, namely, winning region computation, and realizability (see Section~\ref{sec:games}). 


\subsection{Winning region computation}

Columns~\textbf{G-S} to~\textbf{OTF} in Table~\ref{table: results} indicate the running times of different variants of our approach, in seconds, for winning region computation (i.e., when an initial set  of states is not given). The variant \textbf{G-S} is applicable when the given game is a simple game, and it involves no  automaton  construction or product-game formation  (see Section~\ref{sec:simple-games}). Variant~\textbf{GF-P} (resp. \textbf{FG-P}),  involves product constructions with property automata, and is applicable either when the given game is simple or when $\apsi$ (resp.  $\anpsi$)  is deterministic (see Section~\ref{sec:deterministic-games}).  The~\textbf{OTF} variant (see Section~\ref{sec:otf}) is applicable in all cases, as it is the most general.  Any entry \textbf{T/O} in the table denotes a timeout,  of 15 minutes while ``N/A'' indicates not-applicable. 

We observe that when the game is simple,  computing the winning region is  fastest using simple game fixpoint approaches (Variant~G-S).  If both automatons are deterministic,  then the FG-P computation is faster than the GF-P computation.  This is because the former does not require a nested loop, as compared to the latter. The~OTF approach is slower than the other approaches in most of the cases  due to the cost of determinization, but is the only approach that was applicable in one of the benchmarks (Cinderella $C = 1.4$ with a non-simple temporal property). OTF took 7.7 seconds in this case, and returned a non-empty under-approximation of the winning region. The $k$ parameter value given to~OTF is indicated in Column~\textbf{K}.

Our approach is very efficient as per our evaluations. Only on one of the benchmarks did none of the variants terminate within the timeout (Repair-Critical, with non-simple temporal property). On each of the remaining benchmarks, at least one variant of our approach terminated within 43 seconds at  most. 

The other approaches ConSynth and Raboniel are only applicable when an initial set of states is given, and not for general winning region computation.

\subsection{Realizability}

Recall that in this problem, a set of \emph{initial states} is given in addition to the temporal property, with the aim being to check if the chosen player wins from every state in this set. The last three columns in Table~\ref{table: results} pertain to this discussion. Column~\textbf{G} indicates the running time of the \emph{most suitable} variant of our approach for the corresponding benchmark; what we mean by this, Variant~\textbf{G-S} whenever it is applicable, else \textbf{FG-P} if it is applicable, else~\textbf{GF-P} if it is applicable, otherwise~\textbf{OTF}. 

Column~\textbf{C} indicates ConSynth's running times, for the benchmarks for which results were available in other papers. The rows where we show ConSynth's results in red color are ones where we are unsure of its soundness; this is because ConSynth does not determinize non-deterministic automatons, whereas in the literature it has been recognized that in general determinization is required for synthesis~\cite{jobstmann2018graph}.

Column~\textbf{R} indicates Raboniel's running times, obtained from our own runs of their tool. We were not able to encode three benchmarks into Raboniel's system due to the higher complexity of manually encoding these benchmarks; in these cases we have indicated  dashes in the corresponding rows.

It is observable that our approach is much more efficient than the two baseline approaches. We terminate within the given timeout on one all but one benchmark, whereas ConSynth times out on three benchmarks and Raboniel on eight. Considering the benchmarks where both our approach and Raboniel terminate, our approach is \textbf{46x}  faster on average (arithmetic mean of speedups). Considering the benchmarks where both our approach and ConSynth terminate,  our approach is \textbf{244x} faster on average.

A case-by-case analysis reveals that we scale in the challenging Cinderella case where the bucket size $C$ is  $1.9(20)$ (i.e.,  9 repeated 20 times).  We also scale gracefully in the simple elevator examples (Simple-3 to Simple-10), as the number of floors increases from~3 to~10, as compared to Raboniel.   We solve the Cinderella benchmark for $C = 1.4$ with the general LTL specification in 301 seconds (using OTF, with $k = 1$),  which is another challenging case. Raboniel times out for this case.


A detailed list of the specifications used is given in Appendix.  \ref{sec:appendix-specification-table}.

\begin{small}
\begin{table*}[t]
\caption{ Comparison of all approaches of \gensysltl\ with ConSynth and Raboniel. Times are in seconds.  \textbf{T/O} denotes a timeout after 15 minutes.  Abbreviations: \textbf{P} for Player,  \textbf{S} for Specification,  \textbf{DB} for Deterministic B\"uchi,  \textbf{DCB} for Deterministic Co-B\"uchi,  \textbf{G-S} for GenSys-Simple Game,  \textbf{GF-P} for Product B\"uchi Game, \textbf{FG-P} for Product Co-B\"uchi Game,  \textbf{OTF} for On-The-Fly approach, \textbf{K} for OTF bound for which solution was found,  \textbf{C} for ConSynth,  \textbf{R} for Raboniel,  \textbf{G} for GenSys-LTL}
\label{table: results}
    \centering
    \begin{tabular}{lrrrrrr|rrrrr|rrr}
    \toprule
        \textbf{Game} & \textbf{Type} & \textbf{P} & \textbf{S} & \textbf{$|$V$|$} & \textbf{DB?($|$Q$|$)} & \textbf{DCB?($|$Q$|$)} & \textbf{G-S} & \textbf{GF-P} & \textbf{FG-P} & \textbf{OTF} & \textbf{K} & \textbf{C} & \textbf{R} & \textbf{G} \\ 
        \midrule
        Cinderella ($C = 2$) & Real & C & G & 5 & Y (2) & Y (2) & 0.4 & 2.4 & 0.8 & 0.7 & 0 & \textbf{T/O} & \textbf{T/O} & 0.4 \\ 
        Cinderella ($C = 3$) & Real & C & G & 5 & Y (2) & Y (2) & 0.3 & 2.8 & 0.7 & 0.7 & 0 & 765.3 & \textbf{T/O} & 0.3 \\ 
        Repair-Lock & Int & C & G & 3 & Y (2) & Y (2) & 0.3 & 1.0 & 0.4 & 0.4 & 0 & 2.5 & 3.1 & 0.3 \\ 
        Repair-Critical & Int & C & G & 8 & Y (2) & Y (2) & 29.0 & 666.0 & 29.5 & 123.0 & 0 & 19.5 & - & 29.0 \\ 
        Synth-Synchronization & Int & C & G & 7 & Y (2) & Y (2) & 0.3 & 0.6 & 0.3 & 0.4 & 0 & 10.0 & - & 0.3 \\ 
        Cinderella ($C = 1.4$) & Real & E & F & 5 & Y (2) & Y (2) & 0.3 & 1.0 & 0.3 & 2.7 & 3 & 18.0 & \textbf{T/O} & 0.3 \\ 
        Cinderella ($C = 1.4$) & Real & C & GF & 5 & Y (2) &\textbf{N (3)} & 43.0 & 130.0 & N/A & 101.0 & 1 & \textcolor{red}{436.0} &\textbf{ T/O} & 43.0 \\ 
        Cinderella ($C = 1.4$) & Real & C & Gen & 5 & \textbf{N (7)} & \textbf{N (5)} & N/A & N/A & N/A & 7.7 & 0 & \textcolor{red}{4.7} & \textbf{T/O} & 301.0  \\ 
        Cinderella ($C = 1.9(20)$) & Real & C & G & 5 & Y (2) & Y (2) & 42.0 & \textbf{T/O} & \textbf{T/O} & \textbf{T/O} & \textbf{T/O} & - & \textbf{T/O} & 42.0 \\ 
        Repair-Critical & Int & C & Gen & 8 & Y(40) & \textbf{N (6)} & N/A & \textbf{T/O} & N/A & \textbf{T/O} & \textbf{T/O} & \textcolor{red}{53.3} & - & \textbf{T/O} \\ 
        Simple-3 & Int & C & Gen & 1 & Y (5) & \textbf{N (6)} & N/A & 3.3 & N/A & 309.0 & 6 & - & 1.8 & 3.3 \\ 
        Simple-4 & Int & C & Gen & 1 & Y (6) & \textbf{N (7)} & N/A & 4.1 & N/A &\textbf{ T/O} & \textbf{T/O} & - & 2.2 & 4.1 \\ 
        Simple-5 & Int & C & Gen & 1 & Y (7) & \textbf{N (8)} & N/A & 5.8 & N/A & \textbf{T/O} & \textbf{T/O} & - & 5.1 & 5.8 \\ 
        Simple-8 & Int & C & Gen & 1 & Y(10) & \textbf{N(11)} & N/A & 15.6 & N/A & \textbf{T/O} & \textbf{T/O} & - & 27.4 & 15.6 \\ 
        Simple-10 & Int & C & Gen & 1 & Y(12) & \textbf{N(13)} & N/A & 30.3 & N/A & \textbf{T/O} & \textbf{T/O} & - & 108.0 & 30.3 \\ 
        Watertank-safety & Real & C & G & 2 & Y (2) & Y (2) & 0.3 & 0.6 & 0.3 & 0.3 & 0 & - & 19.4 & 0.3 \\ 
        Watertank-liveness & Real & C & Gen & 1 & Y (3) & \textbf{N (4)} & N/A & 2.5 & N/A & 0.7 & 0 & - & 51.0 & 2.5 \\ 
        Sort-3 & Int & C & FG & 3 & Y (2) & N/A & 1.2 & 1.1 & N/A & 0.3 & 0 & - & 51.0 & 1.1 \\ 
        Sort-4 & Int & C & FG & 4 & Y (2) & N/A & 2.2 & 1.2 & N/A & 0.4 & 0 & - & 650.1 & 1.2 \\ 
        Sort-5 & Int & C & FG & 5 & Y (2) & N/A & 5.2 & 1.2 & N/A & 0.4 & 0 & - & \textbf{T/O} & 1.2 \\ 
        Box & Int & C & G & 2 & Y (2) & Y (2) & 0.3 & 0.6 & 0.3 & 0.4 & 0 & 3.7 & 1.2 & 0.3 \\ 
        Box Limited & Int & C & G & 2 & Y (2) & Y (2) & 0.2 & 0.6 & 0.3 & 0.3 & 0 & 0.4 & 0.3 & 0.2 \\ 
        Diagonal & Int & C & G & 2 & Y (2) & Y (2) & 0.2 & 0.6 & 0.2 & 0.4 & 0 & 1.9 & 6.4 & 0.2 \\ 
        Evasion & Int & C & G & 4 & Y (2) & Y (2) & 0.7 & 1.8 & 0.8 & 0.6 & 0 & 1.5 & 3.4 & 0.7 \\ 
        Follow & Int & C & G & 4 & Y (2) & Y (2) & 0.7 & 1.8 & 0.8 & 0.6 & 0 & \textbf{T/O} & 94.0 & 0.7 \\ 
        Solitary Box & Int & C & G & 2 & Y (2) & Y (2) & 0.3 & 0.5 & 0.2 & 0.4 & 0 & 0.4 & 0.3 & 0.3 \\ 
        Square 5 * 5 & Int & C & G & 2 & Y (2) & Y (2) & 0.3 & 0.6 & 0.3 & 0.4 & 0 & \textbf{T/O} & \textbf{T/O} & 0.3 \\ \bottomrule
    \end{tabular}
\end{table*}

\end{small}

\subsection{Discussion on non-termination} \label{subsec: non-termination}There do exist
specifications  where {\gensysltl} will not terminate.  We share this
issue in common with Raboniel.  Consider the game specification: $V =
\{x\},  \cont(x,x') := x' == x - 1 \vee x' = x +1,  \envi(x,x') := x' ==x,
\init(x) := x \geq 0,  \psi(x) := F (x < 0) $.  This example is
realizable.   However,  {\gensysltl} will not terminate as it will
keep generating predicates $x\leq 1$,   $x \leq 2$,  $x \leq 3$, and so on,
which can never cover the initial region $x \geq 0$.

\section{Related Work}
\label{sec:related-work}


\emph{Explicit-state techniques for finite-state games.}
This line of work goes back to B\"uchi and Landweber \cite{Buchi1969},
who essentially studied finite-state games with a B\"uchi winning
condition, and showed that a player always has a finite-memory strategy (if
she has one at all).
Games with LTL winning conditions, where the players play symbols from
an input/output alphabet respectively, were first studied by Pnueli and
Rosner \cite{PnueliR89} who showed the realizabilty question was
decidable in double exponential time in the size of the LTL
specification.
A recent line of work \cite{KupfermanV05,jobstmann2018graph} proposes a
practically 
efficient solution to these games, which avoids the expensive
determinization step, based on Universal Co-B\"uchi Tree Automata.
\cite{filiot2011antichains} extend this direction
by reducing the problem to solving a series
of safety games, based on a $k$-safety automaton ($k$-UCW) obtained 
from a Universal Co-B\"uchi Word (UCW) automaton.

\emph{Symbolic fixpoint techniques.}
One of the first works to propose a symbolic representation of the
state space in fixpoint approaches was \cite{hoffmanWT1992} in the
setting of discrete-event systems.
More recently Spectra \cite{maoz2021spectra} uses BDDs to represent states
symbolically in finite-state safety games.
For infinite-state systems, \cite{AsarinMP94} proposes a logical
representation of fixpoint algorithms for 
boolean and timed-automata based systems,
while \cite{AlfaroHM01} characterizes
classes of arenas for which fixpoint algorithms for safety/reachability/Buchi
games terminate.
An earlier version of our tool called GenSys \cite{gensys} uses a
symbolic fixpoint approach, 
but is restricted to safety games only. In contrast to
all these works, we target the general class of LTL games.
A recent preprint 
\cite{heim2023solving} uses acceleration-based techniques to alleviate
divergence issues in the fixpoint-based game solving approach.
Their approach can terminate in certain cases where our approach does not terminate, such as the one explained in Sec.~\ref{subsec: non-termination}.
However their technique does not attempt to compute the exact winning
region.

\emph{Symbolic CEGAR approaches.}
\cite{HenzingerJM03} considers infinite-state LTL games and proposes a
CEGAR-based approach for 
realizability and synthesis.
Several recent works consider games specified in Temporal Stream Logic
(TSL).
\cite{Finkbeiner0PS19} considers uninterpreted functions and
predicates and convert the game to a bounded LTL synthesis
problem, refining by adding new constraints
to rule out spurious environment strategies.
\cite{raboniel,FinkbeinerHP22,temos} consider TSL modulo theories
specifications and give techniques based on converting to an LTL
synthesis problem, using EUF and Presburger logic,
and Sygus based techniques, respectively.
In contrast to our techniques, these techniques are not guaranteed to
compute precise winning regions or to synthesize maximally permissive
controllers.

\emph{Symbolic deductive approaches.}
In \cite{consynth} Beyene et al propose a constraint-based approach
for solving logical LTL games, which
encodes a strategy as a solution to a system of extended Horn
Constraints. 
The work relies on user-given templates for the unknown relations.
\cite{FarzanK18} considers reachability games,
and tries to find a strategy by first finding one on a finite unrolling
of the program and then generalizing it.
\cite{KatisFGGBGW18,dtsynth,WuQCRFLDX20} consider safety games, and
try to find strategies using forall-exists solvers, a decision-tree based
learning technique, and enumerative search using a solver,
respectively.
In contrast, our work aims for precise winning regions for general
LTL games.

\section{Conclusion}
\label{sec:conclusion}

In this paper we have shown that symbolic fixpoint techniques are
effective in solving logical games with general LTL specifications.
Going forward, one of the extensions we would like to look at
is strategy extraction for general (non-FND) games.
Here one could use tools like AE-Val \cite{FedyukovichGG19} that
synthesize valid Skolem functions for forall-exists formulas.
A theoretical question that appears to be open is whether the class of
games we consider (with real domains in general) are \emph{determined}
(in that one of the players always has a winning strategy from a given
starting state).


\section*{Acknowledgment}
 \addcontentsline{toc}{section}{Acknowledgment}
The authors would like to thank Rayna Dimitrova and Philippe Heim for their comments on a preprint of our paper, which  helped us improve key parts of our paper.

\bibliographystyle{IEEETran}
\bibliography{sample-base}
\clearpage
\appendix
\subsection{Proof of Theorem~\ref{thm:simple-games}}
\label{sec:appendix-simple-games-proof}

We consider each of the four algorithms in turn.
For Algorithm~\ref{proc:gensys-safety} for \textbf{safety games}, let $W_i$
be the set of states denoted by the formula $W$ in the $i$-th
iteration of the algorithm.
We claim that each $W_i$ is exactly the set of states from which
Player~$C$ has a strategy to keep the play in $X$ for at least $2i$ steps.
We prove this by induction on $i$.
$W_0$ corresponds to the set of states represented by $X$, and hence
is exactly the set of states from where $C$ can keep the play in $X$
for at least 0 steps.
Assuming the claim is true upto $i$, consider $W_{i+1}$.
Let $s \in W_{i+1}$.
Then, by construction, there must be a state $t$ with
$(s,t) \in \Delta_{\cont}$ and for all $s'$: $(t,s') \in \Delta_{\envi}$,
we have $s' \in W_i$.
Thus $C$ has a strategy to keep the game in $X$ for two steps starting
from $s$, and to reach a state $s' \in W_{i}$.
From there
(by induction hypothesis) it has a strategy to keep the play in $X$
for at least $2i$ steps. Thus starting from $s$, $C$ has a strategy to keep the
play in $X$ for at least $2i+2$ steps, and hence we are done.
Conversely, suppose $s$ were a state from where $C$ had a strategy to
keep the play in $X$ for at least $2i+2$ steps.
Consider the strategy tree corresponding to this strategy, rooted at
$s$, for a length of $2i+2$ steps.
Clearly, the states at each each level $2k$ ($k<i$) are such that they
have a strategy to keep the play in $X$ for at least $2k$ steps, and hence (by
induction hypothesis) must belong to $W_k$.
But then $s$ must belong to $W_{i+i}$ (since it has strategy to reach
a $W_i$ state in two steps, safely, from $s$).
This completes the inductive proof of the claim.

Suppose now that the algorithm terminates due to $W_{n} \implies
W_{n+1}$. Since the sequence can easily be seen to be telescoping,
this must mean that $W_{n+1} = W_n$.
We now argue that $W_{n+1}$ is the exact winning region for $C$ in
$\myGame$.
Consider a state $s$ in $W_{n+1}$.
We can construct a winning strategy for $C$ from $s$ as follows: By
construction, there is a $t$ such that $(s,t) \in \Delta_\cont$ such
that $t \in X$, and for all $s'$ such that $(t,s') \in
\Delta_{\envi}$, we have $s'
\in W_n$. The strategy for $C$ is simply to play this $t$ from $s$.
Now all resulting states $s'$ belong to $W_n$. Hence they must also
belong to $W_{n+1}$. By a similar argument, we can play a $t'$ from
$s'$ which brings the play back to $W_n$; and so on.
Thus $s \in \winreg{C}{\myGame}$.
Conversely, suppose $s \in \winreg{C}{\myGame}$.
Let $\sigma$ be a winning strategy for $C$ from $s$. Then $\sigma$ is
also a strategy to keep the play in $X$ for at least $2n+2$ steps.
Hence,  by our earlier claim, $s \in W_{n+1}$.
This completes the proof that Algorithm~\ref{proc:gensys-safety}
computes the exact winning region for $C$ in $\myGame$.

For FND games, the strategy automaton output by the algorithm can be
seen to be a winning strategy from $\winreg{C}{\myGame}$ (equivalently
the fixpoint $W_{n}$).
Firstly, the constraints $U_i$ cover the whole of $W_{n}$ by
construction (i.e.\@ $\bigcup_{i =1}^{k} U_i = W_n$).
Furthermore, the strategy keeps the play in $W_n$ forever; since $W_n
\subseteq X$, this means the play is winning for $C$.

Finally, the output strategy can be seen to be \emph{maximally
  permissive}, since from every state $s$ in $\winreg{C}{\myGame}$, we
offer \emph{all} moves $\cont_i$ which let us keep the game in
$\winreg{C}{\myGame}$.

For Algorithm~\ref{proc:gensys-reachability} for \textbf{reachability games}, let $W_i$
be the set of states denoted by the formula $W$ in the $i$-th
iteration of the algorithm.
We claim that each $W_i$ is exactly the set of states from which
Player~$C$ has a strategy to reach some state in $X$ in at most $2i$ steps.
We prove this by induction on $i$.
$W_0$ corresponds to the set of states represented by $X$, and hence
is exactly the set of states from where $C$ can reach some state in $X$
in at most 0 steps.
Assuming the claim is true upto $i$, consider $W_{i+1}$.
Let $s \in W_{i+1}$.
Then, by construction, there must be a state $t$ with
$(s,t) \in \Delta_{\cont}$ and for all $s'$: $(t,s') \in \Delta_{\envi}$,
we have $s' \in W_i$.
Thus $C$ has a strategy to reach a state $s' \in W_{i}$,  from $s$.
From $s'$
(by induction hypothesis) it has a strategy to reach $X$
in $2i$ steps. Thus starting from $s$, $C$ has a strategy to reach $X$ in $2i+2$ steps, and hence we are done.
Conversely, suppose $s$ were a state from where $C$ had a strategy to
keep the play in $X$ for $2i+2$ steps.
Consider the strategy tree corresponding to this strategy, rooted at
$s$, for a length of $2i+2$ steps.
Clearly, the states at each each level $2k$ ($k<i$) are such that they
have a strategy to reach $X$ in $2k$ steps, and hence (by
induction hypothesis) must belong to $W_k$.
But then $s$ must belong to $W_{i+i}$ (since it has strategy to reach
a $W_i$ state in two steps,  from $s$).
This completes the inductive proof of the claim.

Suppose now that the algorithm terminates due to $W_{n+1} \implies
W_{n}$. Since the sequence can easily be seen to be telescoping,
this must mean that $W_{n+1} = W_n$.
We now argue that $W_{n+1}$ is the exact winning region for $C$ in
$\myGame$.
Consider a state $s$ in $W_{n+1}$.
We can construct a winning strategy for $C$ from $s$ as follows: By
construction, there is a $t$ such that $(s,t) \in \Delta_\cont$ such
that $t \in X$, and for all $s'$ such that $(t,s') \in
\Delta_{\envi}$, we have $s'
\in W_n$. The strategy for $C$ is simply to play this $t$ from $s$.
Now all resulting states $s'$ belong to $W_n$. Hence they must also
belong to $W_{n+1}$. By a similar argument, we can play a $t'$ from
$s'$ which brings the play back to $W_n$; and so on.
Thus $s \in \winreg{C}{\myGame}$.
Conversely, suppose $s \in \winreg{C}{\myGame}$.
Let $\sigma$ be a winning strategy for $C$ from $s$. Then $\sigma$ is
also a strategy to reach $X$ in at most $2n+2$ steps.
Hence,  by our earlier claim, $s \in W_{n+1}$.
This completes the proof that Algorithm~\ref{proc:gensys-reachability}
computes the exact winning region for $C$ in $\myGame$.

For FND games, the strategy automaton output by the algorithm can be
seen to be a winning strategy from $\winreg{C}{\myGame}$ (equivalently
the fixpoint $W_{n}$).
Firstly, the constraints $U_i$ cover the whole of $W_{n}$ by
construction (i.e.\@ $\bigcup_{i =1}^{k} U_i = W_n$).
Furthermore, the strategy ensures that from $W_n$,  there exists a play to reach $X$ in at most $2n$ steps.  This is done by ensuring that from a state $s$ exclusively in $W_{i\leq n}$, (which we compute as $C_i$ in Algorithm \ref{proc:gensys-reachability}), we only pick state $s'$ such that there is a $t$ such that $(s,t) \in \Delta_\cont$ such that $t \in X$, and for all $s'$ such that $(t,s') \in \Delta_{\envi}$, we have $s'
\in W_{i-1}$; this ensures that the play must reach $X$ within $2n$ steps.  Thus, the play is winning for $C$.

However,  it must be noted that the output strategy is not \emph{maximally
  permissive} for reachability games since we do not allow strategies that can possibly keep the game in a given $W_i$ for a bounded number of steps before moving on to $W_{i-1}$.  Nevertheless,  we are \emph{maximally
  permissive} in the sense that we consider all moves that can take us to $W_{i-1}$ from $W_i$,  for a given $i$.
  

Algorithms~\ref{proc:gensys-buchi} and \ref{proc:gensys-cobuchi} for \textbf{b\"uchi games} and \textbf{co-b\"uchi games} respectively are written in the style of \cite{Thomas1995} and the proof of correctness and strategy extraction extends similarly.

\subsection{Proof of Theorem~\ref{thm:deterministic-games}}
\label{sec:appendix-deterministic-games-proof}

\textbf{Proof.}
We argue that $C$ has a winning strategy from $W$ in $\myGame$. 

Let $\sigma'$ be a winning strategy for $C$ in $\myGame \myprod
\A_\psi$ for the region $W'$.
Define a strategy $\sigma$ for $C$ in $\myGame$, as follows: for any
odd-length play $w$ in $\myGame$ starting from $s$, we define
\[
\sigma(w) = \{ s' \ | \ \exists p' \in Q: \ (s',p') \in \sigma'(w,\finrun(\A_\psi,w)) \},
\]
where $\finrun(\A_\psi,w)$ is the (unique) run of $\A_\psi$ on $w$.
For convenience, we have represented a sequence of pairs
$(s_0,p_0) (s_1,p_1)\cdots (s_n,p_n)$ as the pair of sequences
$(s_0s_1\cdots s_n, p_0p_1\cdots p_n)$.
It is easy to see that $\sigma$ is a valid strategy from $W$.

Now let $\pi$ be a play in $\myGame$ according to $\sigma$ starting
from $s \in W$.
Then it is easy to see that $\pi' = (\pi, \finrun(\A_\psi, \pi))$ is play
according to $\sigma'$ in $\myGame \myprod \A_\psi$, starting from
$(s,q_0)$ in $W'$.
Since this play is winning for $C$,
$\finrun(\A_\psi, \pi)$ must visit $F$ infinitely often.
Hence $\pi \in L(\A_\psi)$ and consequently $\pi \models \psi$.

Conversely, suppose $C$ has a winning strategy $\sigma$ from a state
$s$ in $\myGame$.
Define a strategy $\sigma'$ for $C$ in $\myGame \myprod \A_\psi$
starting from $(s,q_0)$ as follows.
For every odd-length play $w\cdot s$ according to $\sigma$ starting
from $s$ in $\myGame$, define:
\[
\sigma'(w \cdot s, r\cdot p) = \sigma(w\cdot s) \times \Delta_\T(p,s),
\]
where $\finrun(\A_\psi,w\cdot s) = r \cdot p$.
We can check that $\sigma'$ is a valid strategy.

We now claim that $\sigma'$ is winning for $C$ from $(s,q_0)$ in
$\myGame \myprod \A_\psi$.
Consider a play $(\pi,\rho)$ in $\myGame \myprod \A_\psi$ starting
from $(s,q_0)$. Then it follows that (a) $\pi$ is a play according to
$\sigma$ in $\myGame$ starting from $s$, and (b) $\rho =
\finrun(A_\psi, \pi)$.
By (a) $\pi$ is winning for $C$, and hence $\pi \models \psi$, and
hence, by (b), $\rho$ must visit $F$ infinitely often. This means that
$(\pi,\rho)$ is winning for $C$ in $\myGame \myprod \A_\psi$.

This completes the proof that $W$ is the exact winning region for $C$
in $\myGame$.

\paragraph{Strategy Construction}
Let $\myGame$ be a finitely non-deterministic game, and 
let $\SA'$ be a finitely represented memoryless strategy for $C$ in
$\myGame \myprod \A_\psi$.
We construct a finitely represented finite-memory strategy $\SA$ for
$C$ in $\myGame$ as follows.
$\SA$ is essentially the product of $\SA'$ and $\A_\psi$.
The label of a state controller state $(u,p)$ in $\SA$ is essentially
the label of the $\SA'$ state $u$, though we take only the game moves
and ignore the automaton move.
\qed

For the case when $\A_{\neg\psi}$ is deterministic, 
we can define $\myGame \myprod \A_{\neg\psi}$ in a similar way, except
that the winning condition is now $\F\Gl(\bigwedge_{p \in F} q \neq p)$.
A similar claim to Thm.~\ref{thm:deterministic-games} holds here as well.

\subsection{Need for Deterministic Automata}
\label{sec:need-determinism}

\begin{figure*}
  \centering
  \begin{subfigure}[b]{0.2\linewidth}
    \centering
	\includegraphics[scale=0.3]{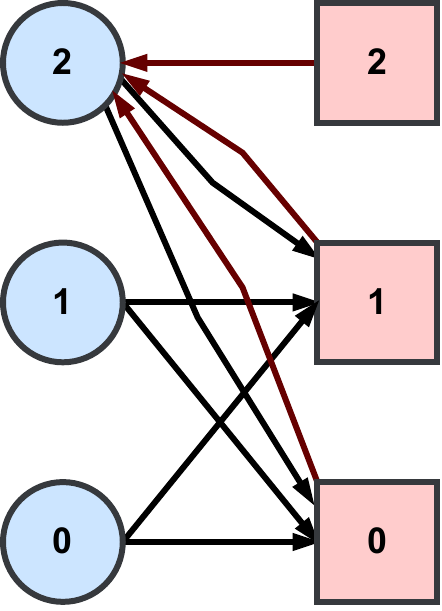}
    \caption{Game graph for $\mathcal{G}_{10}$}
    \label{fig:BA-fin-a-parta}
  \end{subfigure}
  \begin{subfigure}[b]{0.4\linewidth}
    \centering
    \includegraphics[scale=0.4]{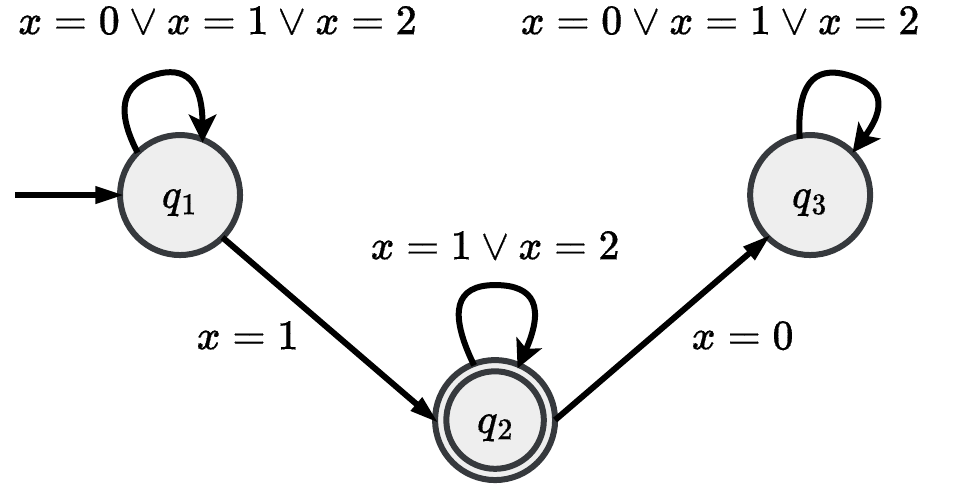}
  \caption{A non-deterministic B\"{u}chi automaton for $\psi = \F(x=1 \myAnd \Gl(\neg (x=0)))$.}
  \label{fig:BA-fin-a-partb}
  \end{subfigure}
  \begin{subfigure}[b]{0.35\linewidth}
    \centering
 \includegraphics[scale=0.3]{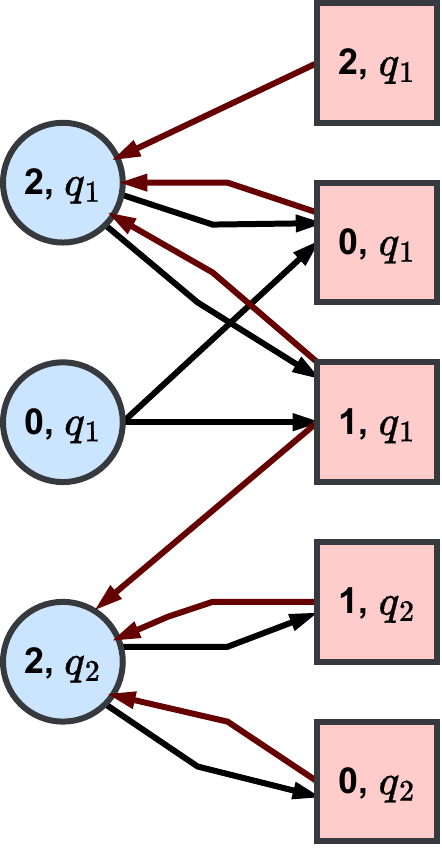}
  \caption{Part of product game graph $\mathcal{G}_{10} \myprod \A_{\psi}$}
  \label{fig:BA-fin-a-partc}
  \end{subfigure}
  \caption{Example illustrating the need for deterministic automata in
    the product game}
  \label{fig:game-non-det-automaton}
\end{figure*}

Consider the LTL game $\mathcal{G}_{10}$ where $V = \{x\}$, $D =
\{0,1,2\}$, $\cont = x' = 0 \myOr x' = 1$, $\envi = x' = 2$, and
\[
\psi = \F(x=1 \myAnd \Gl(\neg (x=0))). 
\]

Fig.~\ref{fig:BA-fin-a-parta} shows the game graph.
A non-deterministic B\"{u}chi automaton for $\psi$ is shown in
Fig.~\ref{fig:BA-fin-a-partb}. Note that there is \emph{no} deterministic
B\"uchi automaton for $\psi$.
We note tha Player~$C$ has a strategy in the game from $x=0$, by
simplying always playing $x=1$. 

Fig.~\ref{fig:BA-fin-a-partc} shows the relevant part of the product
game graph using the non-deterministic automaton for $\psi$.
Note that Player~$C$ does \emph{not} have a winning strategy in the
product game starting from the product game state $(0, q_1)$,
since $E$ can choose to visit state $q_1$ of the
automaton in game state $x=1$, resolving the non-determinism to its
advantage.
Hence a standard computation of the winning region for Player~$C$ in
the product game will \emph{not} contain the state $(0.q_1)$.

This illustrates why we need to have deterministic automata before
constructing the product game.

\subsection{Expressiveness of Logical LTL games}
\label{sec:appendix-expressivity}

Existing specification languages such as Temporal Stream Logic Modulo Theories (TSL-MT) \cite{FinkbeinerHP22} can be modeled using the game-based semantics we consider.  For example, the specification we considered for non-termination in Section \ref{subsec: non-termination}, when specified using TSL-MT is: \[ 0 \leq x  \rightarrow (F(x = 0) \wedge G ([x \leftarrow x + 1] \vee [x \leftarrow x - 1]))\]

In general,  we consider TSL-MT benchmarks where the TSL-MT specifications can be separated into the form $G [move_1] \vee [move_2] \vee ... [move_n]$ and a specification without such moves.  For example,  consider the elevator example encoding from Raboniel \cite{raboniel}:

  \[ 
  \begin{array}{c}
  
  G ([x \leftarrow x -1] \vee [x \leftarrow x +1] \vee [x \leftarrow x]) \ \wedge \\ 
   G (x \geq 1 \wedge x \leq 3) \wedge \\ G( F (x == 1) \wedge F (x == 2) \wedge F (x == 3))
   \end{array}
   \]

The above TSL-MT specification can be separated into two parts,  first, which includes the system (controller + environment) moves, and second, which is the specification.  In this case,  the controller move would be $ x' == x + 1 \vee  x' == x - 1 \vee x' == x $,  the environment moves will be $x' == x$ and the specification would be $G (x \geq 1 \wedge x \leq 3) \wedge G( F (x == 1) \wedge F (x == 2) \wedge F (x == 3))$.

Now consider the TSL-MT specification for the simple elevator given in the TeMos \cite{temos} tool :

\[ G \neg (gt \ b \ t) \rightarrow [up \leftarrow True()]; \]
  \[G (gt\  b \ t) \rightarrow [up \leftarrow False()];\]

The specification is over the input variables $b$, $t$, and the controllable variable $up$. The specification states that the controller must set the variable $up$ to $True$ if $b \leq t$ and $False$ otherwise. In this case,  the updates (or moves) occur within the temporal logic formula. We currently do not consider such specifications; we can support such specifications with minor extensions to our approach. At this point,  GenSys allows the user to separate the system encoding from the specification,  which is easier to express and reason about in our experience. Thus, we use benchmarks from \cite{raboniel} and not \cite{temos}.



Similarly,  we can encode Lustre specifications, which are assume-guarantee-based languages, into logical LTL games. For example, the Cinderella encodings in Lustre format are in the tool JSyn-VG \cite{KatisFGGBGW18}. One can compare the Cinderella benchmarks in \gensysltl\ with JSyn-VG to understand the translation.

\begin{figure}
  \centering
  \includegraphics[scale=0.4]{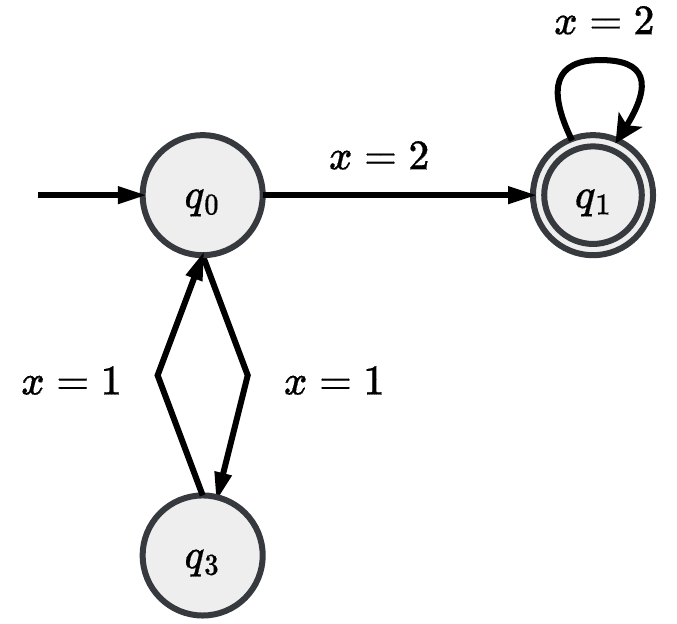}
  \caption{A non-deterministic B\"uchi Automaton which cannot be expressed as an LTL formula}
  \label{fig:nonltl}
\end{figure}

Lastly,  our approach in general takes as input a non-deterministic B\"uchi automaton which is more expressive than an LTL specification.  Thus,  we can express all $\omega$ regular specifications.  An example of a non-deterministic B\"uchi automaton specification which cannot be expressed as an LTL formula is shown in Figure \ref{fig:nonltl}.

\subsection{List of Specifications used in Table \ref{table: results}}
\label{sec:appendix-specification-table}

Table \ref{table: results} describes the specifications used in Table \ref{table:specs}.  Columns 1 - 7 in Table \ref{table:specs}  are the same as Table \ref{table: results}.  Column \textbf{V} shows the set of game variables used in the benchmark.  The column \textbf{LTL Specification} describes the Linear Temporal Logic specification used in the benchmark,  which uses the variables in \textbf{V}.  The last column describes the initial region of the game.  This column is optional; when not provided,  \gensysltl\ computes the winning region for the specification.  The moves of the controller and environment are not present in Table \ref{table:specs} for brevity.  However,  we encourage the reader to visit the respective sources \cite{consynth} \cite{dtsynth} \cite{raboniel} as well as our tool on GitHub\footnote{\url{https://github.com/stanlysamuel/gensys/tree/gensys-ltl}} to view the encodings.  We also describe the conversion of TSL-MT specifications into logical LTL games in Appendix \ref{sec:appendix-expressivity}.

\onecolumn
{ \tiny
    \centering
        \topcaption{ Detailed Specification List used for the evaluation of \gensysltl\ in Table \ref{table: results}.}
    \begin{supertabular}{lllllllp{1in}p{1.5in} p{1in}}

    \toprule
    
        \textbf{Game} & \textbf{Type} & \textbf{P} & \textbf{S} & \textbf{$|$V$|$} & \textbf{DB?($|$Q$|$)} & \textbf{DCB?($|$Q$|$)} & \textbf{V} & \textbf{LTL Specification} & \textbf{Init (when used)} \\ 
        \midrule
        Cinderella ($C = 2$) & Real & C & G & 5 & Y (2) & Y (2) & $\{b1, b2, b3, b4, b5\}$ & $ G (b1 \leq 3.0 \wedge b2 \leq 3.0 \wedge b3 \leq 3.0 \wedge b4 \leq 3.0 \wedge b5 \leq 3.0 \wedge b1 \geq 0.0 \wedge b2 \geq 0.0 \wedge b3 \geq 0.0 \wedge b4 \geq 0.0 \wedge b5 \geq 0.0) $ & $ b1 == 0.0  \wedge b2 == 0.0  \wedge b3 == 0.0  \wedge b4 == 0.0 \wedge b5 == 0.0$ \\ \hline
        Cinderella ($C = 3$) & Real & C & G & 5 & Y (2) & Y (2) & $\{b1, b2, b3, b4, b5\}$ & $ G (b1 \leq 2.0 \wedge b2 \leq 2.0 \wedge b3 \leq 2.0 \wedge b4 \leq 2.0 \wedge b5 \leq 2.0 \wedge b1 \geq 0.0 \wedge b2 \geq 0.0 \wedge b3 \geq 0.0 \wedge b4 \geq 0.0 \wedge b5 \geq 0.0) $ & $ b1 == 0.0  \wedge b2 == 0.0  \wedge b3 == 0.0  \wedge b4 == 0.0 \wedge b5 == 0.0$ \\ \hline
        Repair-Lock & Int & C & G & 3 & Y (2) & Y (2) & $\{pc, l, gl\}$ & $ G \neg ( (pc==2 \wedge l == 1) \vee (pc==5 \wedge l == 0) )$ & $ gl == 0 \wedge l == 0 \wedge pc == 0 $ \\ \hline
        Repair-Critical & Int & C & G & 8 & Y (2) & Y (2) & $ \{ f1a, f1b, t1b, f2a,  \newline f2b,  t2b,  pc1, pc2 \} $ & $ G   \neg (pc1 == 4 \wedge pc2 == 4) \wedge  \neg (pc1 == 8 \wedge pc2 == 7) $ & $ pc1 == 1  \wedge pc2==1 \wedge f1a==0 \wedge  f1b==0 \wedge  t1b==0 \wedge  f2a==0 \wedge  f2b==0 \wedge t2b==0 $ \\ \hline
        Synth-Synchronization & Int & C & G & 7 & Y (2) & Y (2) & $\{ x, y1, y2, z, pc1,  \newline pc2, pc3\}$ & $ G ( \neg ( pc3 == 3 \wedge y1 == y2))$ & $ x == 0 \wedge y1 == 0 \wedge y2 == 0 \wedge z == 0 \wedge pc1 == 1 \wedge pc2 == 1 \wedge pc3 == 1 $ \\ \hline
        Cinderella ($C = 1.4$) & Real & E & F & 5 & Y (2) & Y (2) & $\{b1, b2, b3, b4, b5\}$ & $ F \neg (b1 \leq 1.4 \wedge b2 \leq 1.4 \wedge b3 \leq 1.4 \wedge b4 \leq 1.4 \wedge b5 \leq 1.4 \wedge b1 \geq 0.0 \wedge b2 \geq 0.0 \wedge b3 \geq 0.0 \wedge b4 \geq 0.0 \wedge b5 \geq 0.0) $ & $ b1 == 0.0  \wedge b2 == 0.0  \wedge b3 == 0.0  \wedge b4 == 0.0 \wedge b5 == 0.0$ \\ \hline
        Cinderella ($C = 1.4$) & Real & C & GF & 5 & Y (2) & \textbf{N (3)} & $\{b1, b2, b3, b4, b5\}$ & $ GF (b1 \leq 1.4 \wedge b2 \leq 1.4 \wedge b3 \leq 1.4 \wedge b4 \leq 1.4 \wedge b5 \leq 1.4 \wedge b1 \geq 0.0 \wedge b2 \geq 0.0 \wedge b3 \geq 0.0 \wedge b4 \geq 0.0 \wedge b5 \geq 0.0) $ & $ b1 == 0.0  \wedge b2 == 0.0  \wedge b3 == 0.0  \wedge b4 == 0.0 \wedge b5 == 0.0$ \\ \hline
        Cinderella ($C = 1.4$) & Real & C & Gen & 5 & \textbf{N (7)} & \textbf{N (5)} & $\{b1, b2, b3, b4, b5\}$ & $ GF (b1 \leq  1.4 \wedge b2 \leq 1.4 \wedge b3 \leq 1.4 \wedge b4 \leq 1.4 \wedge b5 \leq 1.4 \wedge b1 \geq 0.0 \wedge b2 \geq 0.0 \wedge b3 \geq 0.0 \wedge b4 \geq 0.0 \wedge b5 \geq 0.0) \vee (GF (b1 \leq  1.4 \wedge b2 > 1.4) \wedge \neg GF (b1 \leq  1.4 \wedge b2 \leq 1.4 \wedge b3 \leq 1.4 \wedge b4 \leq 1.4 \wedge b5 \leq 1.4 \wedge b1 \geq 0.0 \wedge b2 \geq 0.0 \wedge b3 \geq 0.0 \wedge b4 \geq 0.0 \wedge b5 \geq 0.0) \wedge \neg GF (b1 > 1.4)) $ & $ b1 == 0.0  \wedge b2 == 0.0  \wedge b3 == 0.0  \wedge b4 == 0.0 \wedge b5 == 0.0$ \\ \hline
        Cinderella ($C = 1.9(20)$) & Real & C & G & 5 & Y (2) & Y (2) & $\{b1, b2, b3, b4, b5\}$ & $ G (b1 \leq 1.9(20) \wedge b2 \leq 1.9(20) \wedge b3 \leq 1.9(20) \wedge b4 \leq 1.9(20)\wedge b5 \leq 1.9(20) \wedge b1 \geq 0.0 \wedge b2 \geq 0.0 \wedge b3 \geq 0.0 \wedge b4 \geq 0.0 \wedge b5 \geq 0.0) $ & $ b1 == 0.0  \wedge b2 == 0.0  \wedge b3 == 0.0  \wedge b4 == 0.0 \wedge b5 == 0.0$ \\ \hline
        Repair-Critical & Int & C & Gen & 8 & Y(40) & \textbf{N (6)} & $\{ f1a, f1b, t1b, f2a,  \newline f2b, t2b, pc1, pc2\}$ & $ G(pc1 == 3 \implies F pc1 == 4) \wedge G(pc1 == 7 \implies F pc1 == 8) \wedge G(pc2 = 3 \implies F pc2 == 4) \wedge G(pc2 == 6 \implies F pc2 == 7) $ & $ pc1==1  \wedge pc2==1 \wedge f1a==0 \wedge  f1b==0 \wedge  t1b==0 \wedge  f2a==0 \wedge  f2b==0 \wedge t2b==0 $ \\ \hline
        Simple-3 & Int & C & Gen & 1 & Y (5) & \textbf{N (6)} & $\{ x\}$ & $G(1 \leq x \leq 3) \wedge G(F(x==1) \wedge F(x==2) \wedge F(x==3))$ & $ x == 0 $ \\ \hline
        Simple-4 & Int & C & Gen & 1 & Y (6) & \textbf{N (7)} & $\{ x\}$ & $G(1 \leq x \leq 4) \wedge G(F(x==1) \wedge F(x==2) \wedge F(x==3) \wedge F(x==4))$ & $ x == 0 $ \\ \hline
        Simple-5 & Int & C & Gen & 1 & Y (7) & \textbf{N (8)} & $\{ x\}$ & $G(1 \leq x \leq 5) \wedge G(F(x==1) \wedge F(x==2) \wedge F(x==3) \wedge F(x==4) \wedge F(x==5))$ & $ x == 0 $ \\ \hline
        Simple-8 & Int & C & Gen & 1 & Y(10) & \textbf{N(11)} & $\{ x\}$ & $G(1 \leq x \leq 8) \wedge G(F(x==1) \wedge F(x==2)  \wedge ... \wedge F(x==8))$ & $ x == 0 $ \\ \hline
        Simple-10 & Int & C & Gen & 1 & Y(12) & \textbf{N(13)} & $\{ x\}$ & $G(1 \leq x \leq 10) \wedge G(F(x==1) \wedge F(x==2)  \wedge ... \wedge F(x==10))$ & $ x \geq 0 \wedge x \leq 8$ \\ \hline
        Watertank-safety & Real & C & G & 2 & Y (2) & Y (2) & $\{ x1, x2\}$ & $ G  (x1 \geq 0.1 \wedge x1 < 0.7 \wedge x2 \geq 0.1 \wedge x2 < 0.7)$ & $x1 \geq 0.2  \wedge x1 < 0.7  \wedge x2 \geq 0.2 \wedge x2 < 0.7$ \\ \hline
        Watertank-liveness & Real & C & Gen & 1 & Y (3) & \textbf{N (4)} & $\{ x\}$ & $ G (x \geq 0.0 \wedge x < 0.7) \wedge G( x < 0.1 \implies F x \geq 0.4) $ & $ x\geq 0.0 \wedge x < 0.7 $ \\ \hline
        Sort-3 & Int & C & FG & 3 & Y (2) & N/A & $\{a, b, c\}$ & $ FG (a>=b \wedge b>=c) $ & Not used \\ \hline
        Sort-4 & Int & C & FG & 4 & Y (2) & N/A & $\{a, b, c, d\}$ & $ FG (a>=b \wedge b>=c \wedge c>=d) $ & Not used \\ \hline
        Sort-5 & Int & C & FG & 5 & Y (2) & N/A & $\{a, b, c, d, e\}$ & $ FG (a>=b \wedge b>=c \wedge c>=d \wedge d>=e) $ & Not used \\ \hline
        Box & Int & C & G & 2 & Y (2) & Y (2) & $\{ x, y\}$ & $ G (x \leq 3, x \geq 0)$ & Not used \\ \hline
        Box Limited & Int & C & G & 2 & Y (2) & Y (2) & $\{ x, y\}$ & $ G (x \leq 3, x \geq 0)$ & Not used \\ \hline
        Diagonal & Int & C & G & 2 & Y (2) & Y (2) & $\{ x, y\}$ & $ G (y \geq x - 2 \wedge y \leq x + 2)$ & Not used \\ \hline
        Evasion & Int & C & G & 4 & Y (2) & Y (2) & $\{x1, y1, x2, y2\}$ & $ G  \neg (x1 == x2 \wedge y1 == y2) $ & Not used \\ \hline
        Follow & Int & C & G & 4 & Y (2) & Y (2) & $\{x1, y1, x2, y2\}$ & $ G ( (x1\geq x2, y1 \geq y2) \implies  (x1-x2) + (y1-y2)\leq 2)  \wedge (x1 \geq x2, y1<y2) \implies  (x1-x2) + (y2-y1)\leq 2) \wedge (x1<x2, y1 \geq y2) \implies (x2-x1) + (y1-y2) \leq 2) \wedge (x1<x2, y1<y2) \implies  (x2-x1) + (y2-y1) \leq 2) $ & Not used \\ \hline
        Solitary Box & Int & C & G & 2 & Y (2) & Y (2) & $\{ x, y\}$ & $ G (x \leq 3, x \geq 0)$ & Not used \\ \hline
        Square 5 * 5 & Int & C & G & 2 & Y (2) & Y (2) & $\{ x, y\}$ & $G (x \leq 5 \wedge x \geq 0 \wedge y \leq 5 \wedge y \geq 0)$ & Not used \\ \bottomrule
    \end{supertabular}
    \label{table:specs}
}
\twocolumn


\end{document}